# 3D Variational Inference-Based Double-Difference Seismic Tomography Method and Application to the SAFOD Site, California

Hao Yang[1], Xin Zhang[2*], Haijiang Zhang[1,3]

1 Laboratory of Seismology and Physics of Earth's Interior, School of Earth and Space Sciences, University of Science and Technology of China, Hefei 230026, China

2 China University of Geosciences, School of Engineering and Technology, Beijing 100083, China

3 Mengcheng National Geophysical Observatory, University of Science and Technology of China, Hefei 230026, China

* Corresponding Author: Xin Zhang, E-mail: xzhang@cugb.edu.cn

**Abstract**

Seismic tomography is a crucial technique used to image subsurface structures at various scales, accomplished by solving a nonlinear and nonunique inverse problem. It is therefore important to quantify velocity model uncertainties for accurate earthquake locations and geological interpretations. Monte Carlo sampling techniques are usually used for this purpose, but those methods are computationally intensive, especially for large datasets or high-dimensional parameter spaces. In comparison, Bayesian variational inference provides a more efficient alternative by delivering probabilistic solutions through optimization. The method has been proven to be efficient in 2D tomographic problems. In this study, we apply variational inference to solve 3D double-difference (DD) seismic tomographic system using both absolute and differential travel time data. Synthetic tests demonstrate that the new method can produce more accurate velocity models than the original DD tomography method by avoiding regularization constraints, and at the same time provides more reliable uncertainty estimates. Compared to traditional checkerboard resolution tests, the resulting uncertainty estimates measure more accurately the reliability of the solution. We further apply the new method to data recorded by a local dense seismic array around the San Andreas Fault Observatory at Depth (SAFOD) site along the San Andreas Fault (SAF) at Parkfield. Similar to other researches, the obtained velocity models show significant velocity contrasts across the fault. More importantly, the new method produces velocity uncertainties of less than 0.34 km/s for $V_p$ and 0.23 km/s for $V_s$. We therefore conclude that variational inference provides a powerful and efficient tool for solving 3D seismic tomographic problems and quantifying model uncertainties.

## 1 Introduction

Seismic tomography is an important technique used to explore subsurface structures. Ray-based travel time tomography is one such method that utilizes body wave arrival times and has been widely applied in various regions because of its stable solution and low computational requirements (Li *et al.*, 2013; Pesicek *et al.*, 2010; Zhang *et al.*, 2004). In seismic tomography the subsurface is usually parameterized in some way, and a physical relationship is defined which predicts data given a particular set of model parameters. Seismic tomography is therefore an inverse problem (Tarantola, 2005).

Double-difference tomography (tomoDD) is a popular approach for travel time inversion. The method enables simultaneous inversion of velocity structure and earthquake source parameters by using both absolute and differential arrival times (Zhang & Thurber, 2003), which offers improved constraints on source region model parameters. The method has been applied on industrial scale (Meng *et al.*, 2019; Westman *et al.*, 2012; Zhang *et al.*, 2009a), regional scale (Dixit *et al.*, 2014; Thurber *et al.*, 2009; Thurber *et al.*, 2006), and global scale (Pesicek *et al.*, 2010).

Due to the nonlinearity of physical relationships, data sparsity, and data noise, seismic tomographic problems are substantially non-unique (Rawlinson & Spakman, 2016): infinite sets of model parameters can fit the data to within their uncertainty. As a result, it becomes challenging to identify the correct solution (Fichtner *et al.*, 2019). Thus, quantifying the uncertainty of the solution is essential for a more accurate evaluation of inversion results.

Seismic tomographic problems are commonly solved using optimization methods (Aki & Lee, 1976; Tarantola, 2005; Tromp *et al.*, 2005). In these methods, one first establishes an approximate linearized physical relationship around a reference model and then seeks an optimal solution through perturbations of this reference model by minimizing the discrepancy between observed and model-predicted data. Since inversion systems are often underdetermined, a common approach is to use regularization constraints to reduce the uncertainty of the problem (El Ghaoui & Lebret, 1997). However, regularization can deteriorate the model resolutions and add subjective contraints to the models. Furthermore, linearized methods cannot accurately estimate the uncertainty of results for nonlinear problems.

To solve an inverse problem, one should thoroughly analyze and describe the model set by calculating the posterior probability density in the model space using Bayes' theorem (Malinverno, 2002). Monte Carlo sampling methods are frequently utilized for this purpose. For example, the Metropolis-Hastings algorithm is one such method and has been applied to many problems in geophysics (Andrieu & Thoms, 2008), such as seismic travel time tomography (Piana Agostinetti *et al.*, 2015; Zhang *et al.*, 2020a; Zhang *et al.*, 2020b), surface wave dispersion inversion (Li *et al.*, 2019; Zhang *et al.*, 2018), full waveform inversion (Guo *et al.*, 2020; Zhao & Sen, 2021; Zhu, 2018), electromagnetic inversion (Blatter *et al.*, 2019; Mandolesi *et al.*, 2018; Yao *et al.*, 2023), electrical resistivity inversion (Aleardi *et al.*, 2021; Malinverno, 2002; Ramirez *et al.*, 2005) and gravity inversion (Bosch *et al.*, 2006; Izquierdo *et al.*, 2020; Luo, 2010). However, the method becomes computationally inefficient in high dimensional space because of its poor scaling. Advanced Markov Chain Monte Carlo (McMC) techniques have therefore been introduced to the geophysics problems to investigate the model parameter space more extensively. These methods include stochastic Newton McMC (Martin *et al.*, 2012; Zhao & Sen, 2019), Hamiltonian Monte Carlo (Chen *et al.*, 2014; Fichtner *et al.*, 2019; Girolami & Calderhead, 2011), Langevin Monte Carlo(Chen *et al.*, 2014; Durmus *et al.*, 2019) and reversible jump McMC (Green & Hastie, 2009; Mandolesi *et al.*, 2018). However, these methods require adequate model space sampling to approximate the posterior probability distribution (Neal, 1993; Zhang & Curtis, 2020a). As a result, McMC methods are computationally intractable for high dimensional problems because of the curse of dimensionality (Lomax & Curtis, 2001).

Variational inference provides an alternative approach for solving Bayesian inference problems by seeking an optimal distribution in a predefined family of (simplified) probability distributions that approximates the complicated posterior distribution (Blei *et al.*, 2017). This strategy aims to reduce the discrepancy between the posterior and approximation probability distributions. In some cases, this allows for faster convergence to an approximate solution of the posterior distribution, making it beneficial for dealing with large datasets (Blei & Jordan, 2006; Blei *et al.*, 2017). Furthermore, variational inference can be parallelized at the individual sample level, which makes optimal use of modern high-performance computing facilities. In contrast, traditional McMC approaches are difficult to parallelize because each sample depends on the preceding one. In addition, variational inference can be applied to large datasets by dividing the dataset into random minibatches, whereas the same strategy cannot be used in McMC because minibatches may upset the equilibrium conditions required by standard McMC algorithms (Bishop, 2006).

In variational inference, it is important to select a suitable probability density family for approximation because this choice determines the accuracy of the results and the difficulty of the optimization problem. A good choice should be a family that estimates accurately the posterior distribution, while maintains simplicity of the optimization problem so that it is tractable. Different choices can lead to different inference methods. The mean-field approximation method decomposes the joint probability distribution into a product of mutually independent marginal distributions, and consequently each random variable is independent with other variables. By doing this, the original high-dimensional inference problem is converted into a sequence of low-dimensional inference problems. This method has been applied to invert for geologic facies distributions using seismic data (Nawaz & Curtis, 2018, 2019). Although the method is computationally efficient, it usually requires tailored mathematical inference that is challenging to expand. Kucukelbir *et al.* (2017) proposed a method based on the Gaussian variational family, called automatic differential variance inference (ADVI), which utilizes automatic differentiation techniques for direct gradient computation and has been applied in many areas (Zhang & Chen, 2022; Zhang & Curtis, 2020a).

However, this method can lead to biased posterior probability distributions due to its implicit Gaussian approximation and is not well-suited for accurately handling multi-peak posterior problems.

Stein Variational Gradient Descent (SVGD) is a method that utilizes a set of particles to represent the approximating probability distribution. These particles are incrementally adjusted by minimizing the KL divergence, ensuring that the density of these particles matches the target probability distribution in the final state (Leviyev *et al.*, 2022; Liu, 2017; Liu & Wang, 2016). This method enables effective inference in high-dimensional spaces without sampling or acceptance rate calculations, which is typical in traditional McMC techniques. SVGD has been widely applied in geophysical domains such as seismic tomography (Agata *et al.*, 2023; Zhang & Curtis, 2020a), event location (Smith *et al.*, 2022), hydrogeologic inversion (Ramgraber *et al.*, 2021) and full waveform inversion (Zhang & Curtis, 2020b, 2021; Zhang *et al.*, 2023). However, the method has not been applied to 3D seismic travel time tomography to jointly invert for velocity structure and seismic source parameters.

In this study, we use variational inference to solve 3D seismic tomographic problems in the framework of DD tomography. To address the potential limitations of SVGD in greater dimensions, we utilize the Stochastic SVGD (sSVGD, a variant of the SVGD) approach (Leviyev *et al.*, 2022; Zhang *et al.*, 2023). In Section 2, we present the fundamental principles of variational inference, explain the sSVGD technique, and demonstrate its use in event relocation and DD tomography. In Section 3, we conduct a synthetic test and compare the results with those obtained using tomoDD. In section 4, we demonstrate the variational DD seismic tomography method on a dataset recorded by a local seismic array around the SAFOD site along the San Andreas Fault in Parkfield. The high-resolution $V_p$ and $V_s$ models provide clues for studying the relationship between faulting and seismicity. In addition, the new method provides a quantitative measure of the uncertainty in the velocity which cannot be obtained using the traditional method.

## 2 Methods

### 2.1 Variation inference

In Bayesian inference, the *prior* information is represented as a probability density function (pdf), which is updated by introducing observed data to infer the *posterior* distribution of the model parameters (Box & Tiao, 2011). According to the Bayesian principle, the *posterior* pdf can be expressed as:

$$p(\mathbf{m} \mid \mathbf{d}_{\text{obs}}) = \frac{p(\mathbf{d}_{\text{obs}} \mid \mathbf{m})p(\mathbf{m})}{p(\mathbf{d}_{\text{obs}})}, \tag{1}$$

where $\mathbf{m}$ represents a set of model parameters; $\mathbf{d}_{\text{obs}}$ is observed data; $p(\mathbf{m} \mid \mathbf{d}_{\text{obs}})$ represents the *posterior* pdf; $p(\mathbf{m})$ is *prior* pdf describing data-independent *prior* information; $p(\mathbf{d}_{\text{obs}} \mid \mathbf{m})$ is the *likelihood* function, indicating the probability of observing $\mathbf{d}_{\text{obs}}$ if model is true; $p(\mathbf{d}_{\text{obs}})$ is a normalization factor called the *evidence*. Generally, computing $p(\mathbf{d}_{\text{obs}})$ entails a high-dimensional integration of the model parameters, which makes it complicated to directly compute the *posterior* pdf.

Variational inference approximates the *posterior* pdf by finding an optimal distribution within a predefined family of probability distributions. To achieve this, we first define the family of known probability distributions $Q = \{q(\mathbf{m})\}$, for example, a Gaussian distribution or a combination of Gaussians. The variational approach then aims to minimize the difference between the variational distribution $q(\mathbf{m})$ and the *posterior* pdf $p(\mathbf{m} \mid \mathbf{d}_{\text{obs}})$ to find an optimal distribution $q^*(\mathbf{m})$:

$$q^*(\mathbf{m}) = \arg\min_{q \in Q} \text{KL}\left[q(\mathbf{m}) \parallel p(\mathbf{m} \mid \mathbf{d}_{\text{obs}})\right], \tag{2}$$

where the KL divergence measures the difference between two probability distributions (Kullback & Leibler, 1951) and can be expressed as:

$$\text{KL}[q(\mathbf{m}) \parallel p(\mathbf{m} \mid \mathbf{d}_{\text{obs}})] = \text{E}_q[\log q(\mathbf{m})] - \text{E}_q[\log p(\mathbf{m} \mid \mathbf{d}_{\text{obs}})]. \tag{3}$$

The expectations are calculated for the known distribution $q(\boldsymbol{m})$. It can be shown that KL $[q \parallel p] \geq 0$, and equal to 0 if and only if $q(\mathbf{m}) = p(\mathbf{m} \mid \mathbf{d}_{\text{obs}})$ (Liu, 2017; Zhang & Curtis, 2020a). Thus, minimizing the KL divergence yields a best estimate $q^*(\mathbf{m})$ for $p(\mathbf{m} \mid \mathbf{d}_{\text{obs}})$ in the predefined variational family $Q$. Expanding the *posterior* pdf $p(\mathbf{m} \mid \mathbf{d}_{\text{obs}})$ in equation (3) using Bayes' theorem (eq. 1) yields the following equation:

$$\begin{aligned} \text{KL}[q(\mathbf{m}) \parallel p(\mathbf{m} \mid \mathbf{d}_{\text{obs}})] = &\text{E}_q[\log\, q(\mathbf{m})] - \text{E}_q[\log\, p(\mathbf{m}, \mathbf{d}_{\text{obs}})] \\ &+ \log\, p(\mathbf{d}_{\text{obs}}) \end{aligned} . \quad (4)$$

Computing the *evidence* term log $p(\mathbf{d}_{\text{obs}})$ can be challenging in many cases, especially when dealing with high-dimensional model parameters, because it requires a high-dimensional integration, which can be extremely computationally intensive. By shifting this term to the left side of the equation and reversing the sign, a new equation known as the evidence lower bound (ELBO) is obtained:

$$\begin{aligned} \text{ELBO}\,[q] &= \log\, p(\mathbf{d}_{\text{obs}}) - \text{KL}[q(\mathbf{m}) \parallel p(\mathbf{m} \mid \mathbf{d}_{\text{obs}})] \\ &= \text{E}_q[\log\, p(\mathbf{m}, \mathbf{d}_{\text{obs}})] - \text{E}_q[\log\, q(\mathbf{m})] \end{aligned} . \quad (5)$$

This equation establishes a lower limit for the *evidence* log $p(\mathbf{d}_{\text{obs}})$ because KL-divergence is non-negative. Since the *evidence* term log$p(\mathbf{d}_{\text{obs}})$ is constant for a specific problem, minimizing the KL-divergence is equivalent to maximizing the ELBO. Equation 2 can therefore be rewritten as:

$$q^*(\mathbf{m}) = \arg \max_{q \in Q} \text{ELBO}\,[q(\mathbf{m})] \quad (6)$$

## 2.2 Stein variational gradient descent

Many variational approaches in practice rely on predetermined families or forms of distributions, simplifying problem-solving but limiting flexibility to adapt to various target distributions. Emerging variational approaches based on invertible transformations aim to broaden their applicability. These techniques use reversible transformations to alter the original distribution, allowing the modified distribution to be adjusted to various posterior distributions.

By using the density of a set of particles to represent the approximating distribution, SVGD minimizes the KL divergence between the target distribution and the estimated distribution using gradient descent (Liu, 2017; Liu & Wang, 2016). The method introduces smooth transformation $\mathrm{T}(\mathbf{m}) = \mathbf{m} + \epsilon\boldsymbol{\phi}(\mathbf{m})$, where $\mathbf{m} = [\boldsymbol{m_1}, \ldots, \boldsymbol{m_d}]$, $\boldsymbol{m_i}$ denotes the $i$th parameter, $\boldsymbol{\phi}(\mathbf{m}) = [\phi_1,\ldots,\phi_d]$ represents the smooth transformation vector which describes the direction of the perturbation change. It can be proved that the transformation is invertible if the perturbation $\epsilon$ is sufficiently small (Liu & Wang, 2016). Define $\mathcal{H}$ to denote the Hilbert space of regeneration kernels on an $\mathbf{m}$-domain with a positive definite regeneration kernel $k$ satisfying $k(x, y) = \langle\psi(x), \psi(y)\rangle$, and $\mathcal{H}^d$ as a set of multivalued functions with d values. For any $\boldsymbol{\phi}(\mathbf{m}) = [\phi_1,\ldots,\phi_d] \in \mathcal{H}^d$, there exists $\phi_i \in \mathcal{H}$, where $i \in [1,2,\ldots,d]$.

For $\boldsymbol{\phi} \in \mathcal{H}^d$, we define Stein's operator endowed with $\mathcal{H}^d$ and a probability distribution $p$ as,

$$\mathcal{A}_p\boldsymbol{\phi}(\mathbf{m}) = \nabla_{\mathbf{m}}\log\, p(\mathbf{m})\boldsymbol{\phi}(\mathbf{m})^T + \nabla_{\mathbf{m}}\boldsymbol{\phi}(\mathbf{m}). \tag{7}$$

Assuming that T is reversible, define $q_T(\mathbf{z})$ the density of $z = T(\mathbf{m})$ when $\mathbf{m} \sim q(\mathbf{m})$, we have：

$$\nabla_\epsilon \mathrm{KL}[q_T \parallel p]|_{\varepsilon=0} = -\mathrm{E}_q\left[\mathrm{trace}\left(\mathcal{A}_p\boldsymbol{\phi}(\mathbf{m})\right)\right]. \tag{8}$$

Maximizing the expectation on the right-hand side offers the steepest direction of KL divergence, and consequently the KL divergence can be reduced by iteratively stepping a tiny distance in that direction. The optimal $\boldsymbol{\phi}^*_{q,p}(\mathbf{m})$ that maximizes the expectation in equation (8) can be found using the kernelized Stein discrepancy (Liu & Wang, 2016), which can be expressed as:

$$\boldsymbol{\phi}^*_{q,p} \propto \mathrm{E}_{\{\mathbf{m}' \sim q\}}\left[\mathcal{A}_p k(\mathbf{m}', \mathbf{m})\right]. \tag{9}$$

Given the above equation, the KL divergence is minimized by iteratively applying the transformation $\mathrm{T}(\mathbf{m}) = \mathbf{m} + \epsilon\boldsymbol{\phi}^*(\mathbf{m})$ to the initial set of particles $\{\mathbf{m}_i^{\mathbf{0}}\}$. By repeating this process, a distribution path $\{q_l\}_{l=1}^{n}$ between $q_0$ and $p$ can be constructed:

$$\begin{aligned}\boldsymbol{\phi}_l^*(\mathbf{m}) &= \frac{1}{n}\sum_{i=1}^{n}\left[k(\mathbf{m}_j^l, \mathbf{m})\nabla_{\mathbf{m}_j^l}\log\, p(\mathbf{m}_j^l \mid \mathbf{d}_{\mathrm{obs}}) + \nabla_{\mathbf{m}_j^l}k(\mathbf{m}_j^l, \mathbf{m})\right] \\ \mathbf{m}_i^{l+1} &= \mathbf{m}_i^l + \epsilon^l\boldsymbol{\phi}_l^*(\mathbf{m}_i^l),\end{aligned} \tag{10}$$

where $l$ is the current iteration count, $n$ is the number of particles, and $\epsilon$ is the update step. The transformation T is invertible when $\epsilon$ is sufficiently small as the Jacobi matrix can be approximated as a unitary matrix. The process converges to the *posterior* pdf as the number of particles approaches infinity. In the equation of $\boldsymbol{\phi}_l^*(\mathbf{m})$ the first component is a kernel-weighted average of gradients from all particles which pushes the particles toward a high-probability zone, while the second term prevents the particles from collapsing into a single mode.

For the kernel function $k(x, y)$ , we use the Gaussian kernel, also known as the radial basis function (RBF)

$$k(\mathbf{m}, \mathbf{m}') = \exp\left[-\frac{\|\mathbf{m} - \mathbf{m}'\|^2}{2h^2}\right], \tag{11}$$

where $h$ denotes a scale factor, and is usually taken as $h = med^2/\log n$ where med is the median of distances between all pairs of particles. This choice balances the gradient contribution from each particle $\mathbf{m}_i$ with the influence of all other particles as $\sum_{j\neq i} k(\mathbf{m}_i, \mathbf{m}_j) \approx n\exp\left(-\frac{1}{2h^2}\tilde{d}^2\right) = 1$ (Liu, 2017; Zhang & Curtis, 2020a). As the scale factor $h \to 0$, the algorithm falls into independent gradient ascents that maximize the $\log p$ of each particle. In SVGD, the kernel function choice can influence the algorithm efficiency. In this study we utilize the commonly used radial basis function, however, we note that different problems may benefit from other kernel functions, such as inverse multiquadric kernels, Hessian kernels, and Riemann manifold kernels, etc.

To deal with hard constraints of model parameters that often appear in reality, we follow previous studies (Zhang & Curtis, 2020a) by using an invertible logarithmic transformation to transform the model parameters into an unconstrained space:

$$\begin{aligned} \theta_i &= T(m_i) = \log{(m_i - a_i)} - \log{(b_i - m_i)} \\ m_i &= T^{-1}(\theta_i) = a_i + \frac{(b_i - a_i)}{1 + \exp{(-\theta_i)}} \end{aligned} , \tag{12}$$

where $\mathrm{m}_i$ represents the i[th] parameter of the original model, $\theta_i$ is the transformed parameter in an unconstrained space, and $a_i$ and $b_i$ are the lower and upper bounds of the parameter, respectively. The particles can then be updated using Eq. (10) without taking care of the constraints. The final model parameters can be obtained by transforming the particles back into the original space.

**2.3 Stochastic SVGD**

While SVGD has found applications in numerous fields, it may skew samples and exhibit slow convergence under complex distributions, leading to an underestimation of variance in high-dimensional problems (Ba *et al.*, 2021). The sSVGD method creates unbiased samples by introducing noise to the dynamical process, which turns SVGD into a Markov chain Monte Carlo method and makes the method asymptotically convergent (Leviyev *et al.*, 2022). The inverse problems can therefore be solved with a finite number of particles by accumulating many *posterior* pdf samples after a burn-in phase.

Welling & Teh (2011) proposed an expression for a continuous-time Markov process that converges to a target distribution $p(\mathbf{z})$ based on the Euler-Maruyama discretized form of Langevin dynamics：

$$\mathbf{z}_{t+1} = \mathbf{z}_t + \epsilon_t\left[\left(\mathbf{D}(\mathbf{z}_t) + \mathbf{Q}(\mathbf{z}_t)\right)\nabla\log\, p(\mathbf{z}_t) + \Gamma(\mathbf{z}_t)\right] + N\left(\mathbf{0}, 2\epsilon_t\mathbf{D}(\mathbf{z}_t)\right), \tag{13}$$

where $\mathbf{D}(\mathbf{z}_t)$ is a positive semidefinite diffusion matrix, $\mathbf{Q}(\mathbf{z}_t)$ is a skew-symmetric curl matrix, $\Gamma_i(\mathbf{z}_t) = \sum_{j=1}^{d} \frac{\partial}{\partial \mathbf{z}_j}\left(\mathbf{D}_{ij}(\mathbf{z}) + \mathbf{Q}_{ij}(\mathbf{z})\right)$ is the correction term, and $N\left(\mathbf{0}, 2\epsilon_t\mathbf{D}(\mathbf{z}_t)\right)$ represents a Gaussian distribution. The gradient $\nabla\log p(\mathbf{z}_t)$ can be computed using the entire dataset or uniformly randomly selected mini-batches of the data which produces a stochastic gradient approximation. In either case, the method converges asymptotically to the *posterior* pdf as the update step $\epsilon_t \to 0$, and the number of iterations tends to be infinity.

For the set of particles $\{\mathbf{m}_i\}$ defined in section 2.2, we construct an augmented space $\mathbf{z} = (\mathbf{m}_1, \mathbf{m}_2, \dots, \mathbf{m}_n) \in R^{nd}$ that concatenates the $n$ particles, and use Equation (13) to build an efficient sampler that runs multiple interacting Markov chains in parallel. Define a matrix **K**:

$$\mathbf{K} = \frac{1}{n}\begin{bmatrix} k(\mathbf{m}_1, \mathbf{m}_1)\mathbf{I}_{d\times d} & \cdots & k(\mathbf{m}_1, \mathbf{m}_n)\mathbf{I}_{d\times d} \\ \vdots & \ddots & \vdots \\ k(\mathbf{m}_n, \mathbf{m}_1)\mathbf{I}_{d\times d} & \cdots & k(\mathbf{m}_n, \mathbf{m}_n)\mathbf{I}_{d\times d} \end{bmatrix}, \tag{14}$$

where $k(\mathbf{m}_i, \mathbf{m}_j)$ is the kernel function introduced in section 2.2, and $\mathbf{I}_{d\times d}$ is the unit matrix. According to the characteristics of kernel functions, the matrix **K** is positive definite (Gallego & Insua, 2018). Given this definition, Eq. (10) can be written in matrix form:

$$\mathbf{z}_{t+1} = \mathbf{z}_t + \epsilon_t[\mathbf{K}\nabla\log\, p(\mathbf{z}_t) + \nabla\cdot\mathbf{K}]. \tag{15}$$

This indicates that SVGD method can be treated as a specific form of Eq. (13) with $\mathbf{D}_\mathbf{K} = \mathbf{K}$, $\mathbf{Q}_\mathbf{K} = \mathbf{0}$ and no noise term $N\left(\mathbf{0}, 2\epsilon_t\mathbf{D}(\mathbf{z}_t)\right)$. Hence, by introducing the noise term, we construct a stochastic gradient McMC algorithm with SVGD gradients (Gallego & Insua, 2018), called stochastic SVGD,

$$\mathbf{z}_{t+1} = \mathbf{z}_t + \epsilon_t[\mathbf{K}\nabla \log\, p(\mathbf{z}_t) + \nabla \cdot \mathbf{K}] + N(\mathbf{0}, 2\epsilon_t \mathbf{K}). \tag{16}$$

This process converges asymptotically to $p(\mathbf{z}) = \prod_{i=1}^{n} p(\mathbf{m}_i \mid \mathbf{d}_{\text{obs}})$. Note that from the definition of the kernel function matrix $\mathbf{K}$, the noise term becomes arbitrarily small when the number of particles is sufficiently large. As a result, sSVGD produces the similar results as standard SVGD.

In practice, estimating the *posterior* pdf using Equation 16 requires drawing samples from $N(\mathbf{0}, 2\epsilon_t\mathbf{K})$ . This requires computing the lower triangular Cholesky decomposition of the matrix $\mathbf{K}$, which is often computationally expensive. To efficiently compute the noise term, we define a diagonal matrix consisting of $d$ repeated kernel matrices $\overline{\mathbf{K}}$**,**

$$\mathbf{D}_{\mathbf{K}} = \frac{1}{n}\begin{bmatrix} \overline{\mathbf{K}} & & \\ & \ddots & \\ & & \overline{\mathbf{K}} \end{bmatrix}, \tag{17}$$

where $\overline{\mathbf{K}}$ is a $n \times n$ matrix with $\overline{\mathbf{K}}_{ij} = k(\mathbf{m}_i, \mathbf{m}_j)$. The matrix $\mathbf{D}_{\mathbf{K}}$ can be computed by $\mathbf{D}_{\mathbf{K}} = \mathbf{P}\mathbf{K}\mathbf{P}^{\mathbf{T}}$, where $\mathbf{P}$ is a permutation matrix,

$$\mathbf{P} = \left[\begin{array}{ccccccccc} 1 & & & & & & & & \\ & & 1 & & & & & & \\ & & & & \ddots & & & & \\ & & & & & & 1 & & \\ \hline & 1 & & & & & & & \\ & & & 1 & & & & & \\ & & & & & \ddots & & & \\ & & & & & & & 1 & \\ \hline & \ddots & & \ddots & & \ddots & & \ddots & \\ \hline & & 1 & & & & & & \\ & & & & 1 & & & & \\ & & & & & & \ddots & & \\ & & & & & & & & 1 \end{array}\right]. \tag{18}$$

Applying this permutation matrix to a vector $\mathbf{z}$ rearranges the order of the vectors from the order of listing them in particle order to listing them in the first coordinate of all particles, followed by the second, and so on. We can now create a SVGD method with a noise term in the SG-MCMC framework (Chen *et al.*, 2016; Nemeth & Fearnhead,

2021), using the following equation to produce noise $\boldsymbol{\eta}$,

$$\begin{aligned}\boldsymbol{\eta} &\sim N(\mathbf{0}, 2\epsilon_t \mathbf{K}) \\ &\sim \sqrt{2\epsilon_t}\mathbf{P}^{\mathrm{T}}\mathbf{P}N(\mathbf{0}, \mathbf{K}) \\ &\sim \sqrt{2\epsilon_t}\mathbf{P}^{\mathrm{T}}N(\mathbf{0}, \mathbf{D_K}) \\ &\sim \sqrt{2\epsilon_t}\mathbf{P}^{\mathrm{T}}\mathbf{L}_{\mathbf{D_K}}N(\mathbf{0}, \mathbf{I}),\end{aligned} \tag{19}$$

where $\mathbf{L}_{\mathbf{D_K}}$ is the lower triangular Cholesky decomposition of the matrix $\mathbf{D_K}$. It is easy to obtain $\mathbf{L}_{\mathbf{D_K}}$ because $\mathbf{D_K}$ is a block diagonal matrix. In practice, the number of particles $n$ is usually moderate, and consequently the calculation of the noise term is computationally negligible.

### 2.4 Variational hypocenter inversion

We first apply the sSVGD method to earthquake source parameters inversion. The source parameter $\mathbf{h}$ consists of spatial coordinates $[x, y, z]$ and origin time $\tau_0$. In the framework of Bayesian inversion, $\mathbf{h}$ is inferred by combining a priori information with the observed data,

$$p(\mathbf{h} \mid \mathbf{d}_{\mathrm{obs}}) = \frac{p(\mathbf{d}_{\mathrm{obs}} \mid \mathbf{h})p(\mathbf{h})}{p(\mathbf{d}_{\mathrm{obs}})}, \tag{20}$$

where $p(\mathbf{d}_{\mathrm{obs}} \mid \mathbf{h})$ can be expressed as:

$$p(\mathbf{d}_{\mathrm{obs}} \mid \mathbf{h}) \propto \exp\left(-\frac{1}{2}\sum_i \frac{\left[t_i^{obs} - t_i^{pred}\right]^2}{\sigma_i^2}\right), \tag{21}$$

where $i$ is the data index for arrival time t, $\sigma_i$ is the uncertainty of arrival time $t_i^{obs}$. For a given seismic velocity model, the theoretical arrival time is calculated using the following equation:

$$t^{pred} = \tau_0 + f(x, y, z), \tag{22}$$

where $f$ is a function of the location of an earthquake source.

In reality, earthquakes usually occur close to each other in the space. Seismic waves generated by those earthquakes and recorded at the same station therefore travel along similar ray paths in space far from the source region. As a result, earthquake locations can be more accurately determined using differential arrival times from a pair of events recorded at the same station because biases caused by the velocity structure in space far from the source region cancel out. In addition, differential arrival times can be more accurately determined by using cross correlations of waveforms, which in turn improves accuracy of earthquake locations. This idea has been used in earthquake relocation (Smith *et al.*, 2022; Waldhauser, 2001). In this study, we use differential arrival times in a Bayesian framework, and in such case, the *likelihood* function can be expressed as:

$$p(\mathbf{d}_{\text{obs}}^{dd} \mid \mathbf{h}) \propto \sum_{a} \sum_{b} \frac{1}{\sqrt{\sigma_a^2 + \sigma_b^2}} \exp{(A)}, \tag{23}$$

$$A = -\frac{\left[\left(t_a^{obs} - t_b^{obs}\right) - \left(t_a^{pred} - t_b^{pred}\right)\right]^2}{\sigma_a^2 + \sigma_b^2}, \tag{24}$$

Where subscripts $a$ and $b$ represent different earthquakes, $\sigma$ is the uncertainty of data. Given this equation, the *posterior* pdf can be obtained using the sSVGD algorithm (equation 16).

### 2.5 Variational double-difference tomography

In this section, we further apply sSVGD to 3D seismic travel time tomography. For the likelihood function $p(\mathbf{d}_{\text{obs}} \mid \mathbf{m})$ we use a Gaussian distribution,

$$p(\mathbf{d}_{\text{obs}} \mid \mathbf{m}) \propto \exp\left(-\frac{1}{2}\sum_{obs_i} \frac{\left[t_i^{obs} - t_i^{pred}\right]^2}{\sigma_i^2}\right), \tag{25}$$

where $i$ is the data index, $\sigma_i$ is the uncertainty of data, $t_i^{obs}$ is the observed arrival time, and $t_i^{pred}$ is the model predicted arrival time.

Similar to section 2.4, we introduce differential arrival times to improve accuracy of velocity estimates in the source region. The *likelihood* function $p(\mathbf{d}_{\text{obs}}^{dd} \mid \mathbf{m})$ can be expressed as:

$$p(\mathbf{d}_{\text{obs}}^{dd} \mid \mathbf{m}) \propto \sum_{a} \sum_{b} \frac{1}{\sqrt{\sigma_a^2 + \sigma_b^2}} \exp{(A)}, \tag{26}$$

$$A = -\frac{\left[\left(t_a^{obs} - t_b^{obs}\right) - \left(t_a^{pred} - t_b^{pred}\right)\right]^2}{\sigma_a^2 + \sigma_b^2}, \tag{27}$$

where subscripts $a$ and $b$ represent different earthquakes, $\sigma$ is uncertainty associated with the measurement error, and $\mathbf{d}_{\text{obs}}^{dd}$ represents the differential arrival time data.

The locations of earthquake sources significantly impact the velocity structure estimates (Zhang *et al.*, 2020b). Therefore, we invert for the source location and velocity structure jointly using arrival times. Here, before performing joint inversion, we conduct a separate inversion of source parameters to obtain relatively accurate source locations to enhance the accuracy and efficiency of the joint inversion.

## 3. Synthetic tests

### 3.1 Earthquake Location

We first use sSVGD to perform earthquake location. The procedure of the whole process is illustrated using a toy example in Fig. 1, in which two earthquakes occurred within a horizontally layered velocity model with vertically increasing velocity and are recorded by 54 stations. The distribution of these stations is illustrated in Fig. 2a, which is from the actual station distribution in the Longmenshan Fault Zone (LFZ), China (Liu *et al.*, 2021). The arrival times and their gradients with respect to earthquake locations are calculated using the pseudo-bending approach (Um & Thurber, 1987). Initially, 150 particles are randomly generated from a uniform prior distribution in the 3D space of interest, and updated using equation (15) for 2000 iterations at which point the average misfit value across particles becomes stationary (step 2 and 3 in Fig. 1).

Due to the broad prior information (within a depth range of ±15 km and a horizontal range of ±1° around the true location) and non-uniqueness of the problem, particles may be trapped in local maxima (those dots around the near surface and at greater depths in the rightmost panel in Fig. 1), which is a common problem in nonlinear inversion. We note that this can be reduced by increasing the iterations of sampling.

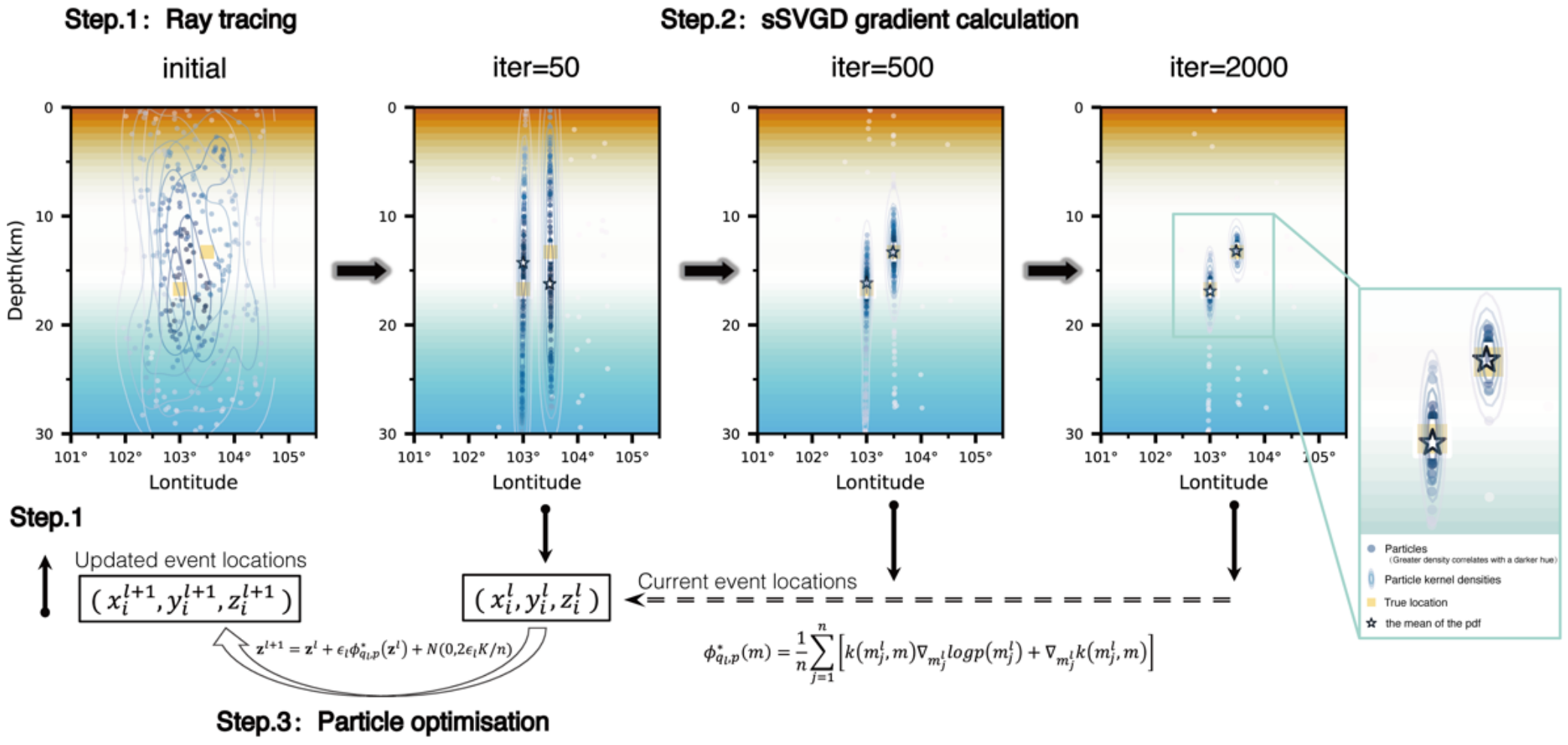

Figure 1. Seismic event relocation process by sSVGD. The yellow squares represent the true source locations, while the blue dots and contour lines indicate the distribution and kernel density of particles, with darker colors representing higher densities. The optimization process of particle positions is shown from left to right. The white stars indicate the locations with the mean of the probability distribution in each iteration.

To further assess the method's performance, we locate 25 seismic events simultaneously using the same station distribution and velocity model (Fig. 2). Prior information for the earthquake location is provided as follows: in both latitude and longitude it has a range of [true location - 30km, true location + 30km], and has a range of [5km, 25km] in depth. We then perform sSVGD for 1000 iterations using 20 particles. The results show that all 25 events can be located with very small uncertainties if the true velocity model is used (Fig. 2d). By contrast, if we assume that the velocity model has an alternating velocity bias of 8% (Fig. 2e), there will be errors in all seismic relocation results, especially in the boundary where the kernel density of source locations have multi-peak distributions. Nonetheless, the method still achieves satisfactory results, particularly in the horizontal direction. This demonstrates that an inaccurate velocity model causes more bias in source depth than in horizontal location. Additionally, we observed that seismic events located at the boundary of the study area exhibit higher uncertainties. The distributions have an elliptical shape, tilting toward the center of the study area. This is primarily because for seismic events at the boundary most of the stations are distributed on one side and the recorded data lack wide azimuthal information, which results in weaker constraints.

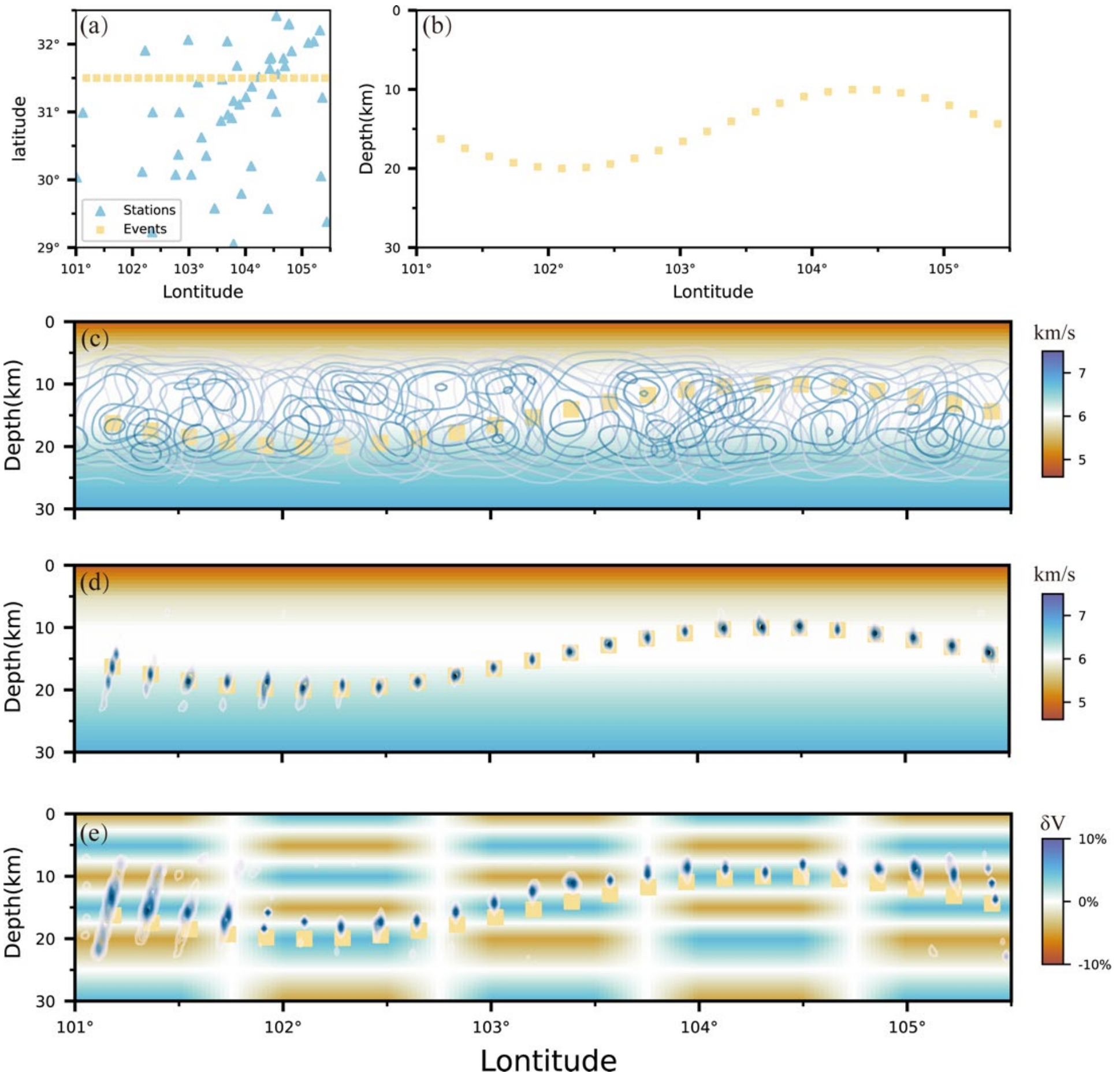

Figure 2. Precision analysis of the sSVGD seismic relocation method. (a) Horizontal distribution of stations (blue triangles) and actual seismic sources (yellow squares). (b) Vertical distribution of actual seismic sources along 31°N latitude. (c) The velocity model and initial distribution of event locations. Blue contour lines indicate kernel density, with darker colors representing higher densities. (d) Relocation results using the accurate velocity model, with the black dots representomg the mean of the probability distribution. (e) Relocation results using a velocity model with errors, applying ± 8% perturbation.

### 3.2 Variational double-difference seismic tomography

We demonstrate the variational DD seismic tomography method by conducting a synthetic test. The test uses a real data distribution comprising 54 stations and 2,275 seismic events (Fig. 3). The true velocity model contains a 10% high-velocity anomaly between 5 km and 10 km depth in the specified location (white box in Fig. 4) within a background velocity model with vertically increasing velocities. The prior information of velocity is set as a uniform distribution with an interval of 1.6 km/s ($V_p$) and 1.2s km/s ( $V_s$ ) at each depth, centered at the background velocity value. The prior information for the earthquake locations is a uniform distribution centered around the true coordinates, with a horizontal interval of 10 km and a vertical interval of 2 km. Arrival times of these events recorded at each station are calculated using the pseudo bending method according to the real earthquake catalog. In addition, Gaussian noise of 0.1 s is added to these data. For sSVGD, 50 particles were generated from the prior information and updated using equation 15 for 500 iterations with an additional burn-in period of 200 iterations.

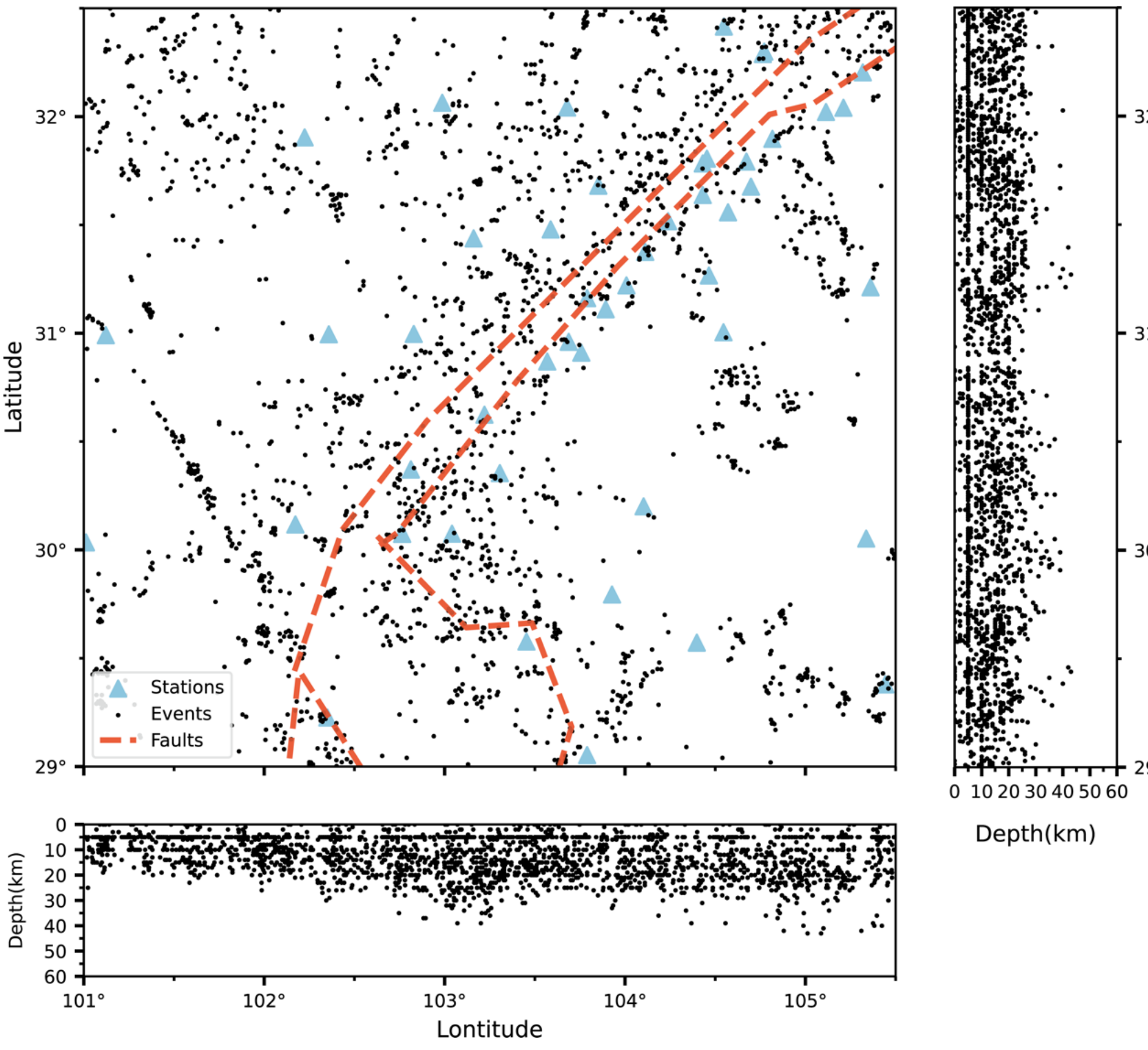

Figure 3. Spatial distribution of 54 seismic stations (blue triangles) and 1932 seismic events (black dots) used for seismic tomography in the synthetic tests. The right panel shows the distribution of events in the latitude-depth section, while the bottom panel shows their distribution in the longitude-depth section. Red dashed lines indicate faults within the region.

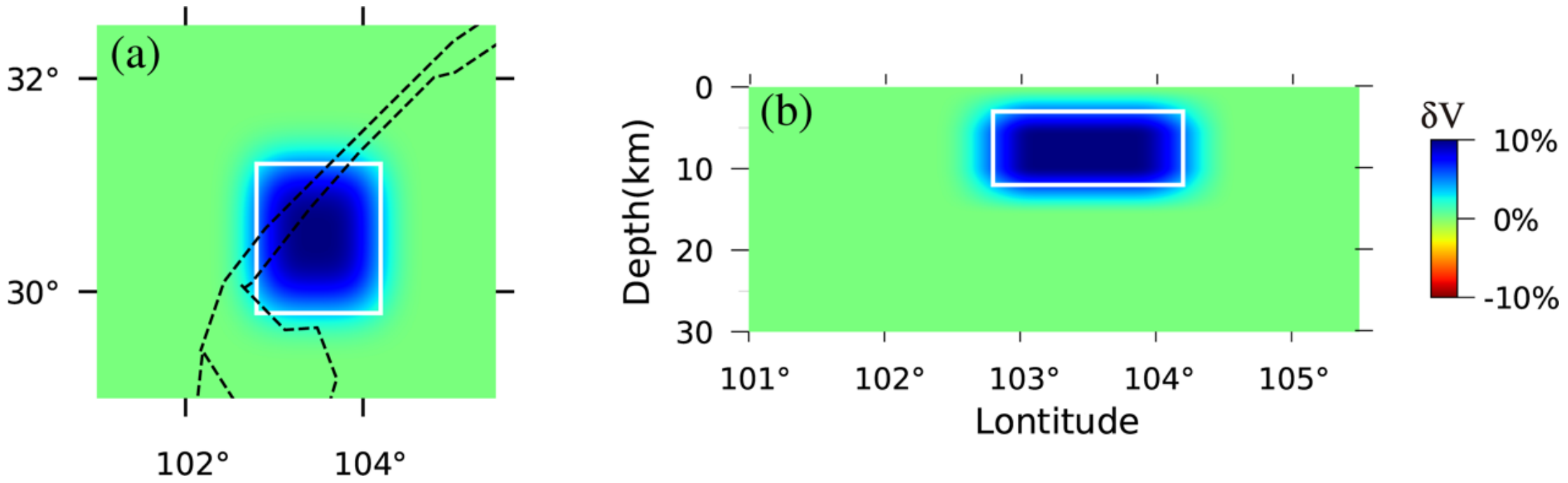

Figure 4. (a) The horizontal section of the true velocity model at depth of 10 km, and (b) the vertical section at latitude = 31° . The white box indicates the location of 10% high-velocity anomaly.

Figs 5 and 6 display horizontal sections of the mean and standard deviation models for $V_p$ and $V_s$ respectively at the depth of 10 km. Overall the mean model captures the true structure, including the high-velocity anomaly in the middle of the study region. The standard deviation model shows high uncertainties in the peripheral regions because of inadequate data coverage. To further analyze the results, we show marginal probability distributions of velocity parameters at three locations: point 1 is within the high velocity anomaly, point 2 is at the boundary between the anomaly and the background velocity, and point 3 is in the region close to the edge of the model. At point 1 and 2 the marginal distributions show high probability values around the actual value. By contrast, at point 3 the results show multi-peak distributions and exhibit higher uncertainty due to inadequate data coverage.

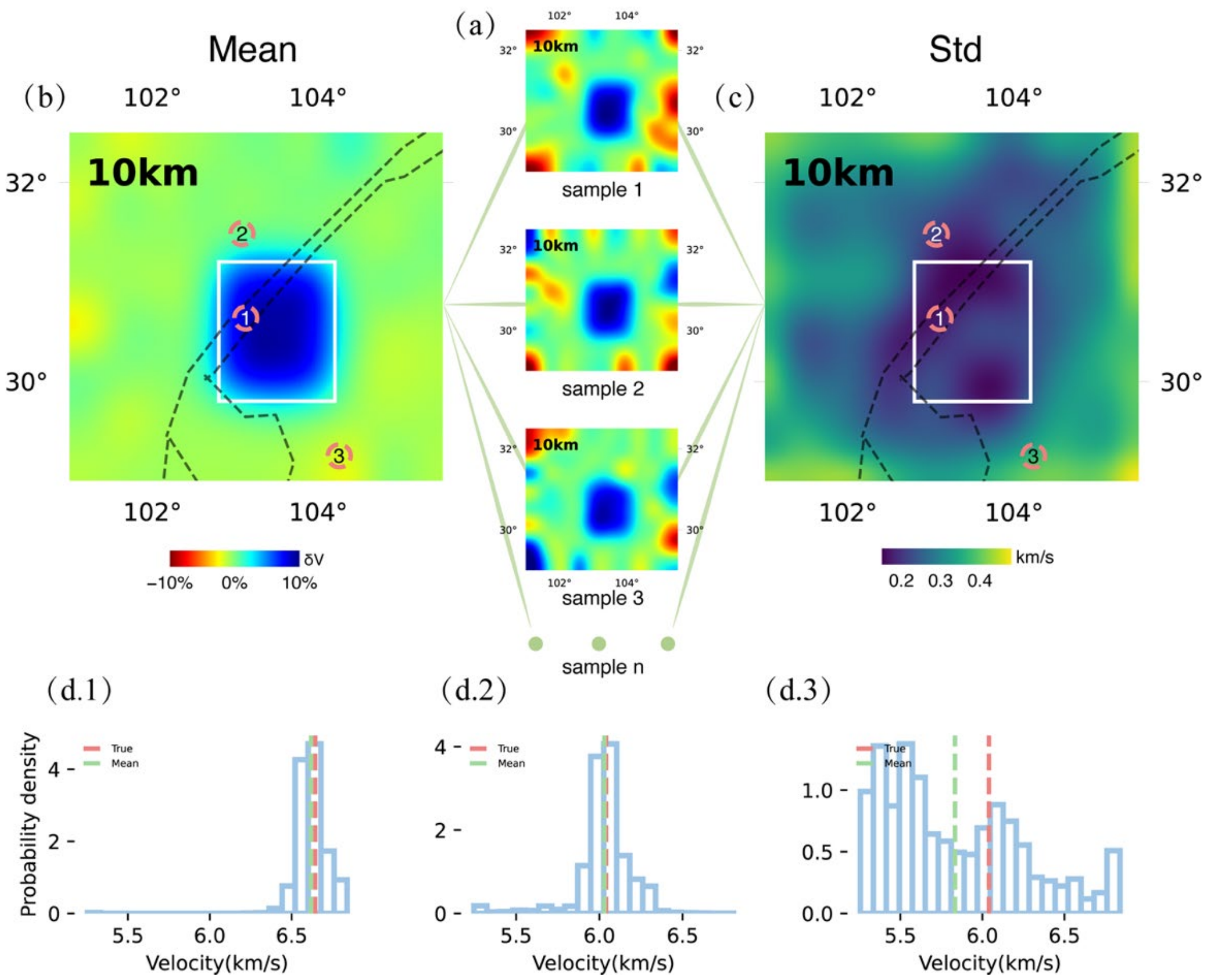

Figure 5. Inversion results of $V_p$ at the horizontal slice at 10km depth by the variational DD seismic tomography method. (a) Examples of three particles randomly selected from the total 50 particles at the last iteration. (b) and (c) show the mean and standard deviation computed using the samples after the burn-in period. (d) Marginal probability distributions of seismic velocities at three locations denoted in subfigures (b) and (c). The green dashed line represents the mean value, and the red dashed line represents the true value.

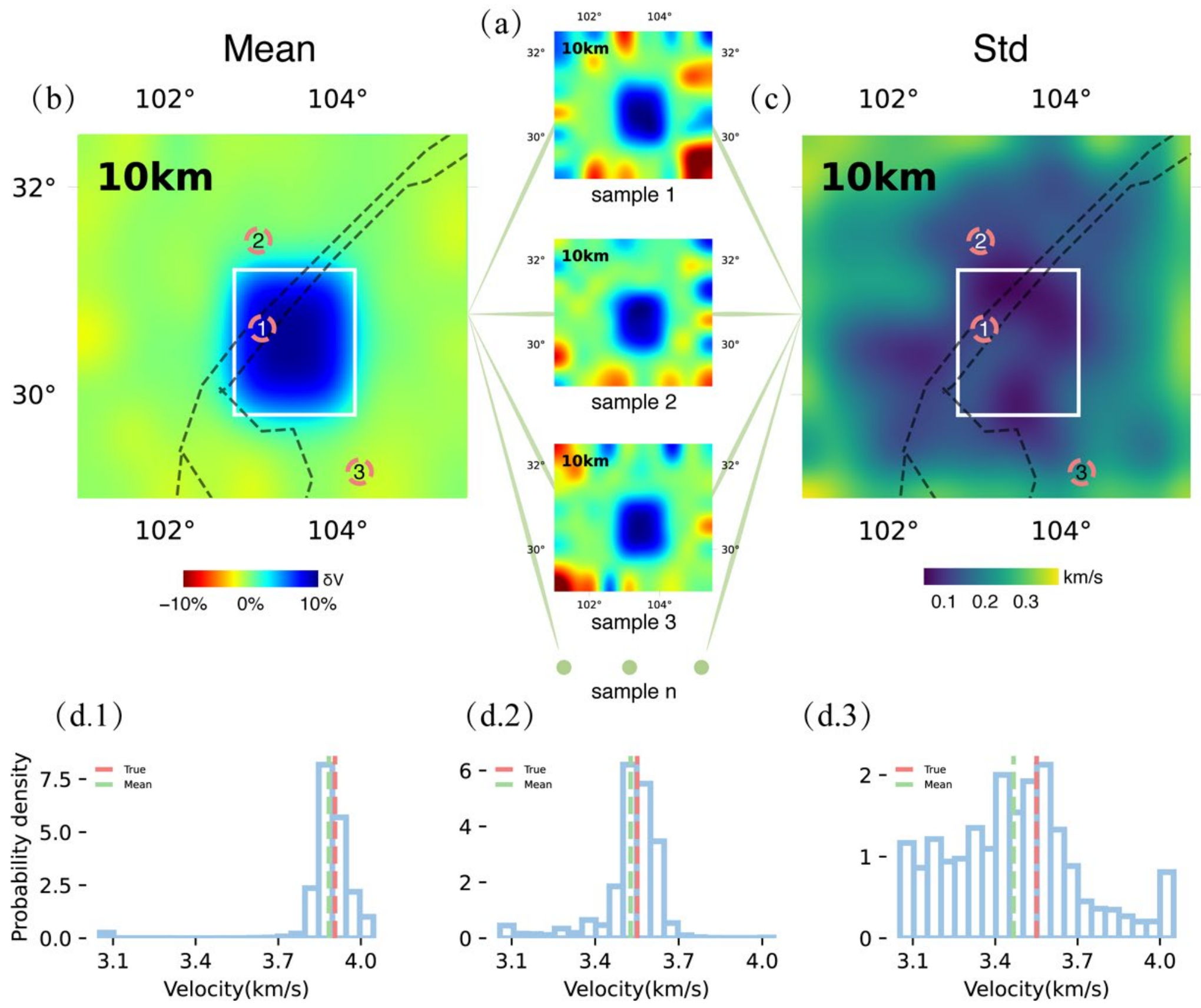

Figure 6. The same as Figure 5 but for $V_S$.

To further assess the performance of our method, we compare the results with those obtained using tomoDD (Zhang & Thurber, 2003). The reliability of the results obtained using tomoDD is assessed using the standard checkerboard test. For the initial velocity model, we use the background model that does not include high-speed anomalies. The regularization parameters (damping and smoothing factors) are selected using the L-Curve method (damping is set to 40 and smoothing to 3) (Hansen, 1999).

The root mean square (RMS) errors decreased from 0.4064s (calculated using the prior mean model) to 0.1747s for the variational DD tomography and to 0.1713s for tomoDD, respectively. Although the average model obtained using variational DD tomography shows slightly lower data fitting compared to that obtained using tomoDD, it is important to note that variational DD tomography does not require an accurate initial model (i.e., strong prior information) as in tomoDD, which is not always available in practice. In fact, the average model obtained using variational DD tomography provides better data fitting in practical scenarios where obtaining accurate initial models is challenging. This will be further discussed in the real application in section 4.

Overall both approaches successfully recover the target anomalies. However, there are artifacts in the tomoDD results at depths of 0 km and 15 km (Figs. 7 and 8). This is because in tomoDD, regularization is selected based on the assumption that the subsurface medium is smooth, which can cause artifacts in regions with strong velocity contrasts, particularly in the depth direction (Fig. 9). Note that this is difficult to avoid by simply reducing the regularization term because small regularization can cause instability in the system. In addition, the checkerboard test shows good resolutions at the depth of 15km, making it difficult to distinguish the anomalies as artifacts. Moreover, since regularization parameters in checkerboard tests typically differ from those utilized in the actual inversion, it becomes even more difficult to interpret the obtained results in practice. By contrast, variational DD tomography does not rely on regularization constraints and thus does not show similar artifacts. Besides, it estimates uncertainty from real data, which is more accurate than those estimated from synthetic data as in the checkerboard test.

The results obtained using variational DD tomography indicate that most parts of the model are accurately reconstructed, despite some errors at 0 km depth and around the boundary region. This likely reflects that sSVGD has not converged sufficiently in those regions because the posterior distribution in these regions have broad extent across the space due to limited data coverage, and consequently the small number of samples (25000) are not sufficient to represent the posterior distribution. This can also be observed by the high uncertainties of $V_p$ and $V_s$ in these regions.

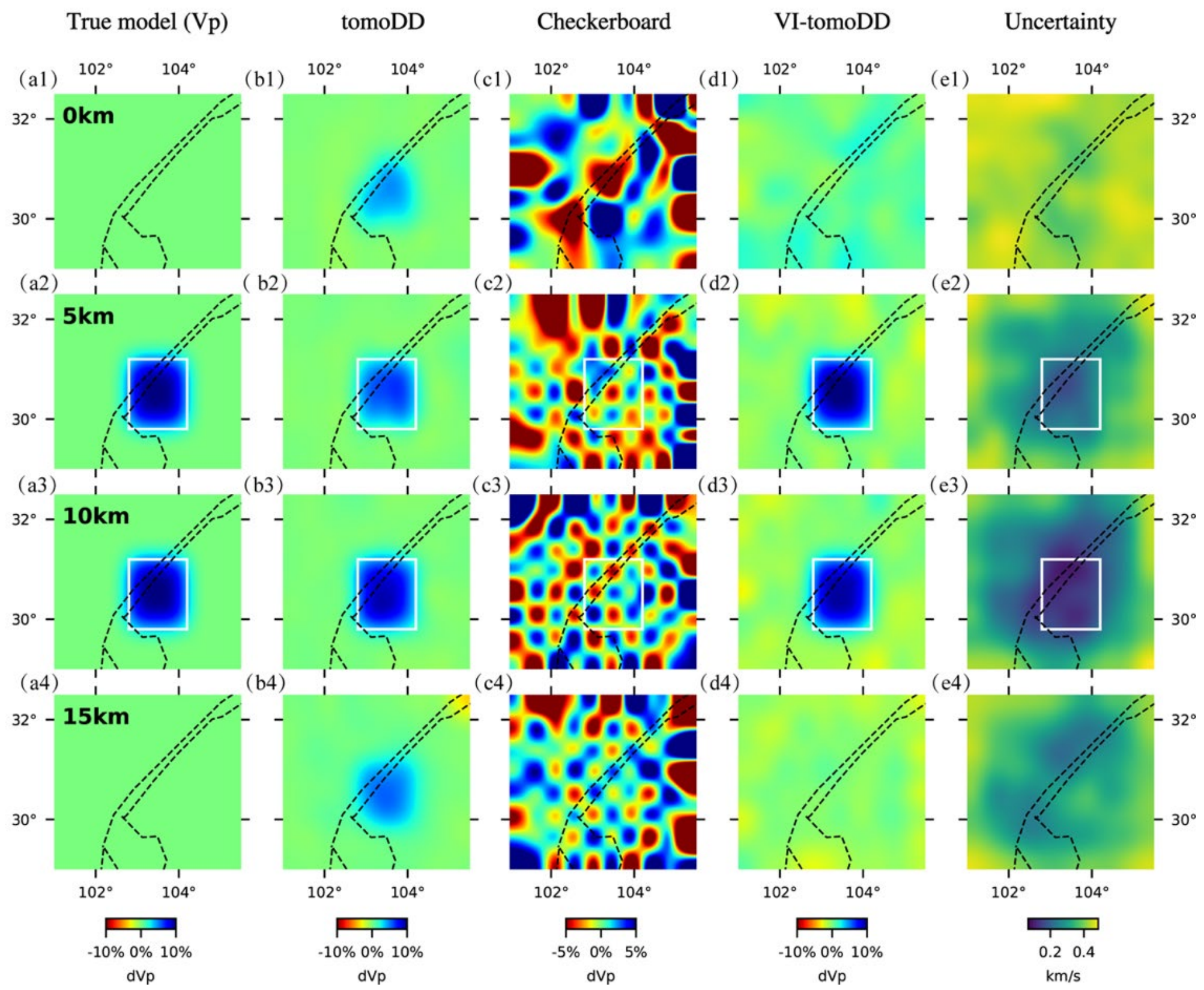

Figure 7. Comparison of $V_p$ results obtained using tomoDD and variational DD tomography. (a) True model; (b) Inversion results obtained using tomodd; (c) Checkerboard test results obtained using tomoDD; (d) The mean model obtained using variational DD tomography; (e) Uncertainties obtained using variational DD tomography. The areas outlined in white boxes indicate the defined high-velocity anomaly regions.

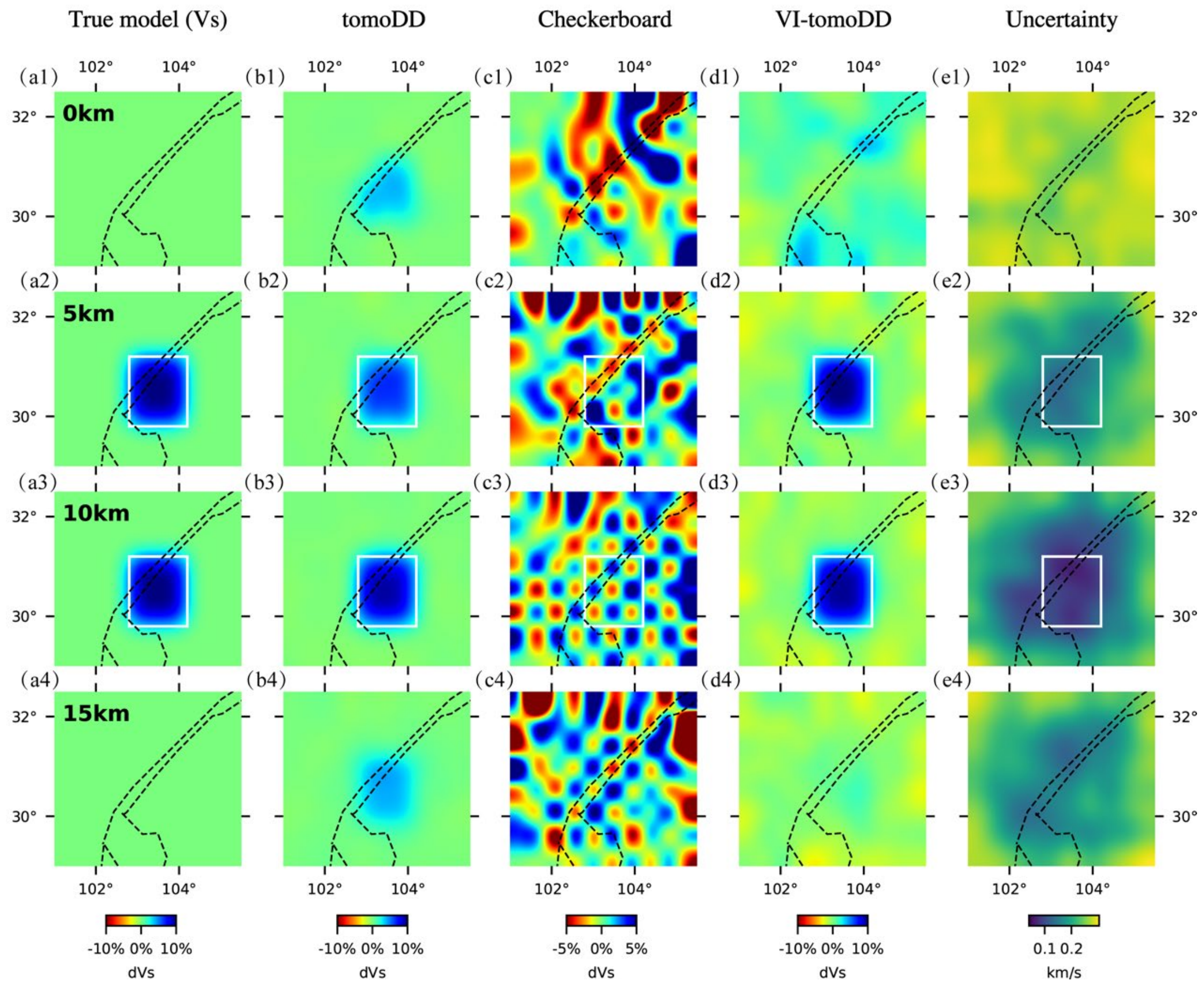

Figure 8. The same as Figure 7 but for the Vs model.

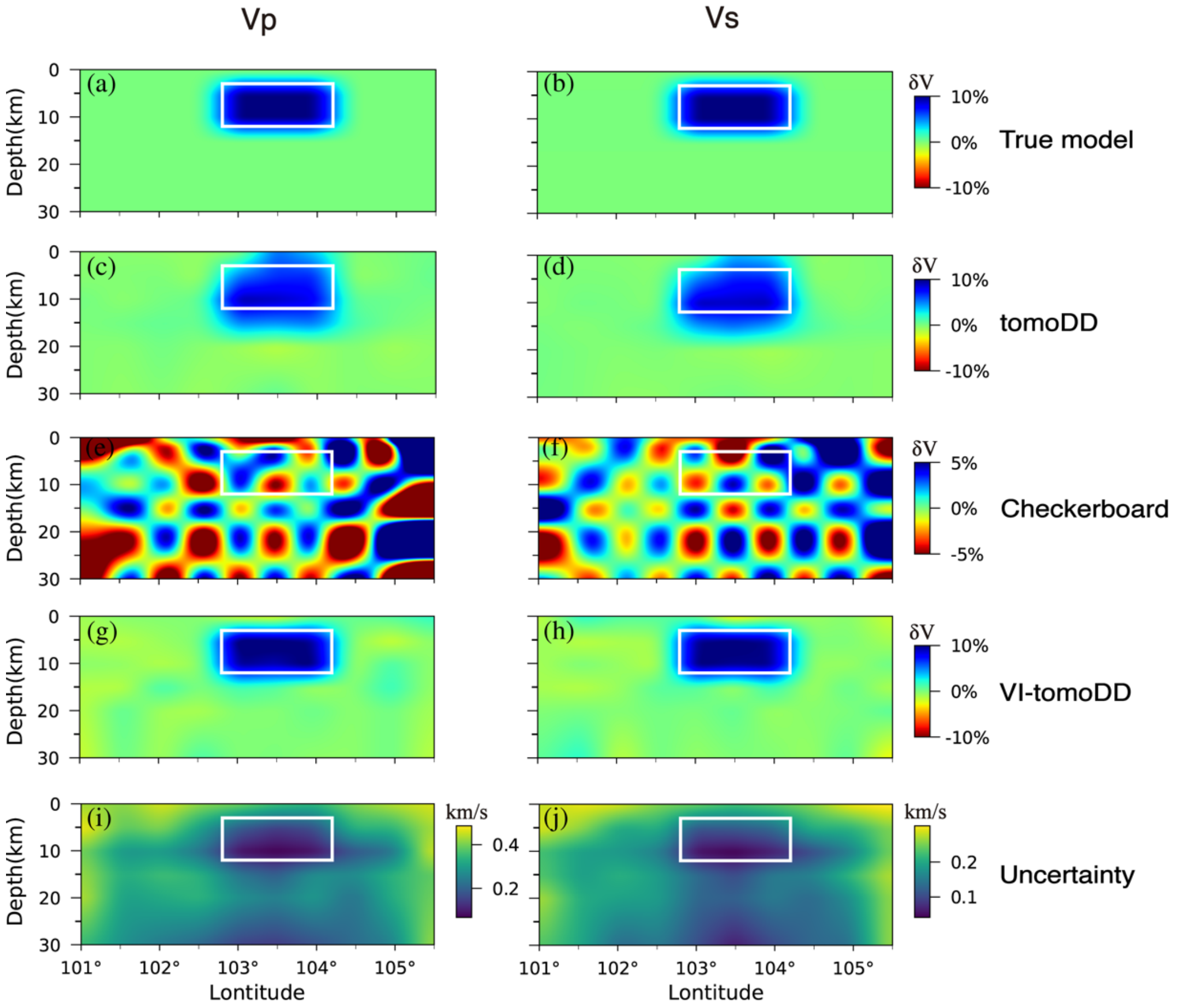

Figure 9. Comparison of results obtained using tomoDD and variational DD tomography along a vertical section at 31° N latitude. (a, b) True model; (c, d) Inversion results obtained using tomoDD; (e, f) Checkerboard test results obtained using tomoDD; (g, h) The mean model obtained using variational DD tomography; (i, j) Uncertainties obtained using variational DD tomography. The areas outlined by in white boxes indicate the defined high-velocity anomaly regions.

## 4 Application to the SAFOD site, Parkfield

In this section, we apply the variational DD tomography method to investigate the upper crustal structure around the SAFOD site of the Parkfield region. In this study, we use the same dataset as that used in (Zhang *et al.*, 2009b) for the direct comparison of variational DD tomography and tomoDD.

The workflow comprises two primary steps: (1) Hypocenter inversion using variational inference; and (2) DD tomography based on variational inference. Event relocations should be prioritized before conducting a joint inversion of seismic sources and velocity structure due to the coupling effect between them.

### 4.1 Data and Inversion Details

The dataset primarily originates from the Parkfield Area Seismic Observatory (PASO), a highly concentrated network consisting of 59 stations strategically deployed around the SAFOD drilling site between 2000 and 2002. Additional data sources include the UC Berkeley High-Resolution Seismic Network, the U.S. Geological Survey CALNET seismic network, 32 three-component sensors installed in the SAFOD pilot hole, data from 17 blast points, and data from the PSINE survey and 1994 blasts. The active source data include 66 blast points detonated in October 2002, recorded by the Parkfield Area Seismic Observatory network and borehole sensors, along with six blasts detonated in 1994 and other supplementary datasets. In total, there are 351,646 absolute P-wave arrival times, 116,934 absolute S-wave arrival times, 77,825 differential P-wave arrival times, and 30,214 differential S-wave arrival times.

The study area is centered on the SAFOD main hole. The Y-axis is rotated 40° counterclockwise from the north, aligning with the local trend of the SAF. Fig. 10 displays the geographical locations of grid points, and the grid nodes in depth are positioned at -0.5, 0.0, 0.5, 1.0, 2.0, 4.0, 7.0, and 10 km, respectively. For the prior information, we use uniform distributions centered on a velocity model with velocity contrasts across the SAF from previous inversion results (Zhang *et al.*, 2009b). The intervals of the uniform distributions are set to 1.2 km/s and 0.8 km/s for $V_p$ and $V_s$ , respectively. Similarly, for seismic source locations, we use uniform prior distributions centered on earthquake catalog locations with an interval of 1 km. Due to a lack of reliable data uncertainty assessment, we use a data noise level of 0.05s for all arrival times, which is estimated from the standard deviation of arrival times from multiple pickings.

We first relocate seismic events using a fixed velocity structure from Zhang *et al.* (2009b). Based on prior information, we generated 80 particles, which are updated using sSVGD for 1500 iterations. The mean locations are obtained by averaging 80,000 samples from the final 1000 iterations. Similarly, for the joint inversion process, we generated 80 particles and performed sSVGD for 1500 iterations. Since the seismic events have already been relocated, we narrowed the prior distribution's width for the source coordinates from 1 km to 0.5 km. Finally, we obtain the mean and standard deviation from 80,000 samples produced during the last 1000 iterations.

In this study, we use the 'posterior standard deviation' of the model parameters to identify data-constrained regions (Piana Agostinetti *et al.*, 2020). Specifically, regions where the standard deviation of $V_p$ and $V_s$ is less than 0.34 km/s and 0.23 km/s, respectively, are considered to have high resolutions. This criterion is based on the principle that when the posterior standard deviation is smaller than the prior standard deviation, it indicates that the data have provided new constraints.

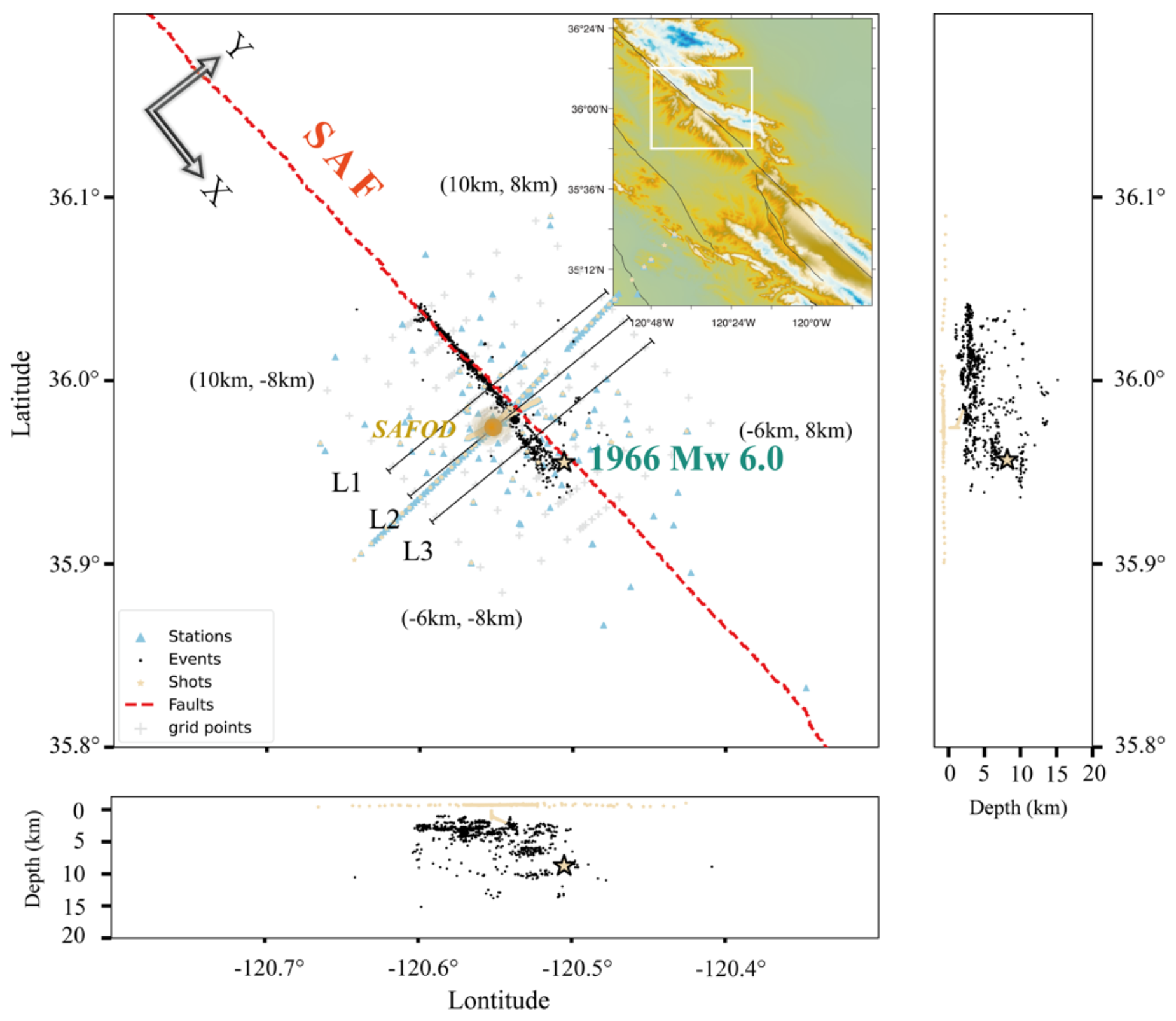

Figure 10. Distribution of events, stations and inversion grid nodes around the SAFOD site. Black dots represent regional seismicity. Blue triangles represent stations, yellow points represent explosions, yellow stars represent the Mw 6.0 1966 epicenter, gray "+" represent grid points, large yellow circle represents SAFOD site, and red dashed lines marks the surface trace of the SAF.

### 4.2 Results

The three-dimensional structure around Parkfield exhibits a significant velocity contrast across the southern portion of the SAF, with higher velocities on the southwest side (see Figs. 11 and 12). This contrast is particularly pronounced in the shallow part (less than 4 km). Directly beneath the surface rupture of the SAF, $V_p$ is less than 2.8 km/s, and $V_s$ is below 1.5 km/s, which may indicate fractured rocks, loose structures, and possibly fluid-filled zones. Earthquakes are concentrated along the transition between high and low velocities. The southeastern segment of the SAF shows less earthquake activity, and the velocity contrast between the two sides is slightly lower than that in the northern segment. This suggests a change in fault rupture behavior. In this region, the reduced creep rate means that stress cannot be relieved through minor earthquakes, leading to several Mw 6 earthquakes over the past centuries (Piana Agostinetti *et al.*, 2020).

In the deeper part (below 4 km), within the southwestern part of the SAF and parallel regions, the $V_p$ is about 7 km/s, while the $V_s$ is approximately 3.5 km/s, indicating a mix of granodiorite and high-velocity rocks. High $V_p$ and relatively lower $V_s$ correspond to the metamorphosed diorite face resulting from the intrusion of Mesozoic granite. Several studies have identified the high-velocity body on the southwest side of the fault as the Salinian block (Zhang *et al.*, 2009b), primarily composed of granite and characterized by relatively high density. On the northeast side, the Franciscan mélange overlies the unmetamorphosed sedimentary rocks of the Great Valley Sequence. The Franciscan mélange primarily consists of chert, greenschist, flint, serpentinite, and blue gneiss (Zhang *et al.*, 2007). These rock types are loosely deposited and exhibit lower wave velocities, which is consistent with our results.

Along the SAF, a narrow $V_p$ low-velocity zone can be observed (see Fig. 12, within the dashed circle of the L3 profile). This zone is approximately 1~2 km wide, with some sections extending to depths of up to 9 km, which is deeper than those reported in previous studies. We also observe that most earthquakes are distributed along the edge of the high velocity body on the southwest side of the fault. This pattern is particularly evident in the L3 profile, where earthquakes at depths of 5 to 8 km are distributed along the contours of the high velocity body's edge. Notably, the 1966 Mw 6.0 earthquake occurred in this vicinity. In the L2 cross-section, between 2.5 and 5.5 km (marked by the solid yellow box), the results show relatively lower $V_p$ and higher $V_s$, and earthquakes are concentrated. This may indicate that these regions are fractured due to fault activity, but the materials are relatively sturdy and retain significant shear resistance. We propose that these regions could have accumulated significant seismic stress, which may indicate potentially high seismic activity and high risk of larger earthquakes.

The white-shaded areas represent regions with high "posterior standard deviations", indicating areas that are not sufficiently constrained by data. In the standard deviation models (Fig. A), we find a narrow zone of high uncertainty near the faults, which represents the uncertainty of the boundary of the velocity contrasts as noted in several previous studies (Galetti *et al.*, 2015; Zhang *et al.*, 2018).

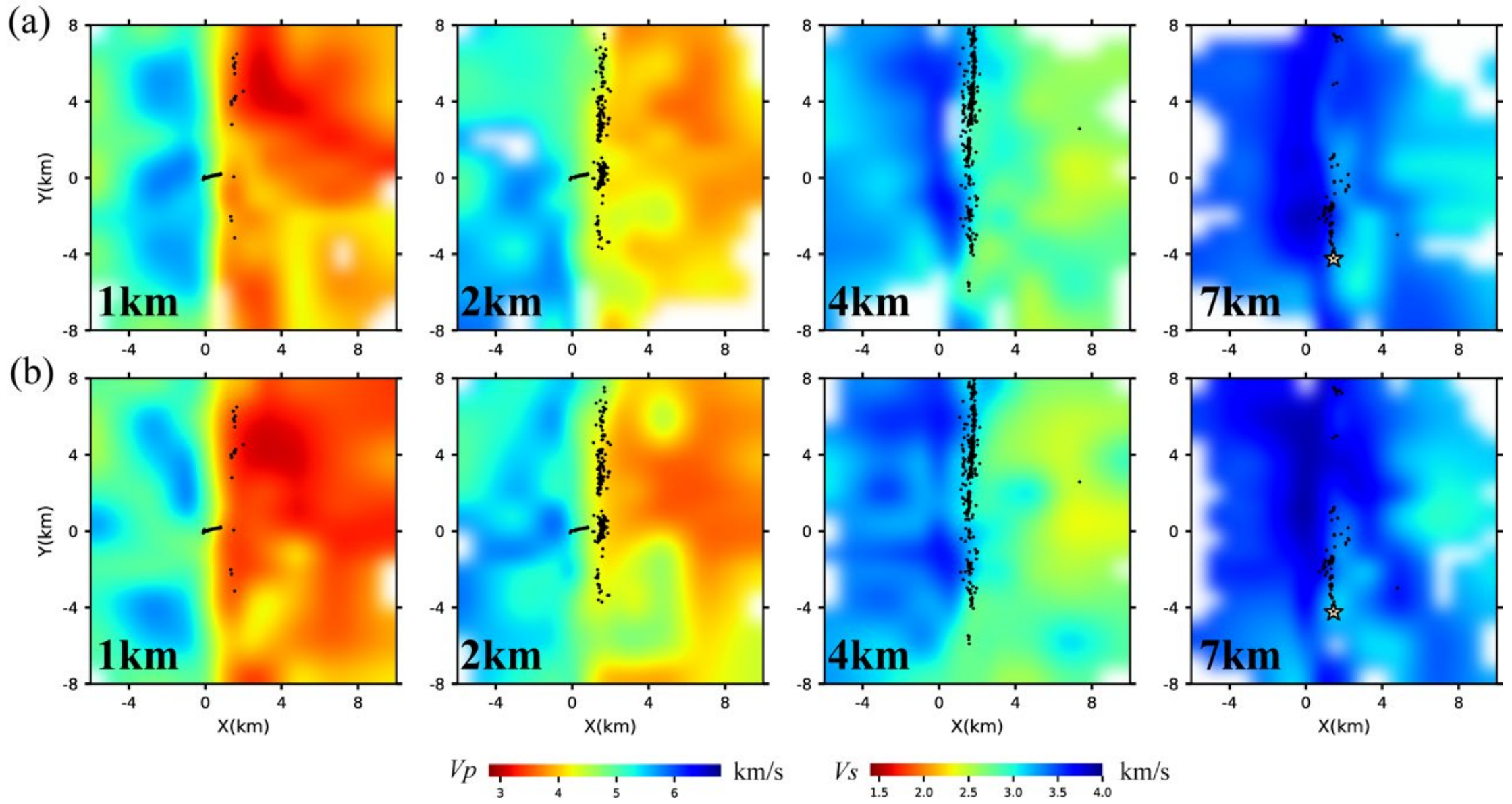

Figure 11. Horizontal slices of the mean Vp (top) and mean Vs (lower panel) models at different depths from variational DD tomography method. Relocated earthquakes (black dots) occurring within ±1 km from the slice are shown, and yellow stars represent the Mw 6.0 1966 epicenter. The white-shaded areas represent regions with high standard deviations ( $V_p$ std > 0.34 and $V_s$ std > 0.23), which indicate poor data constraints.

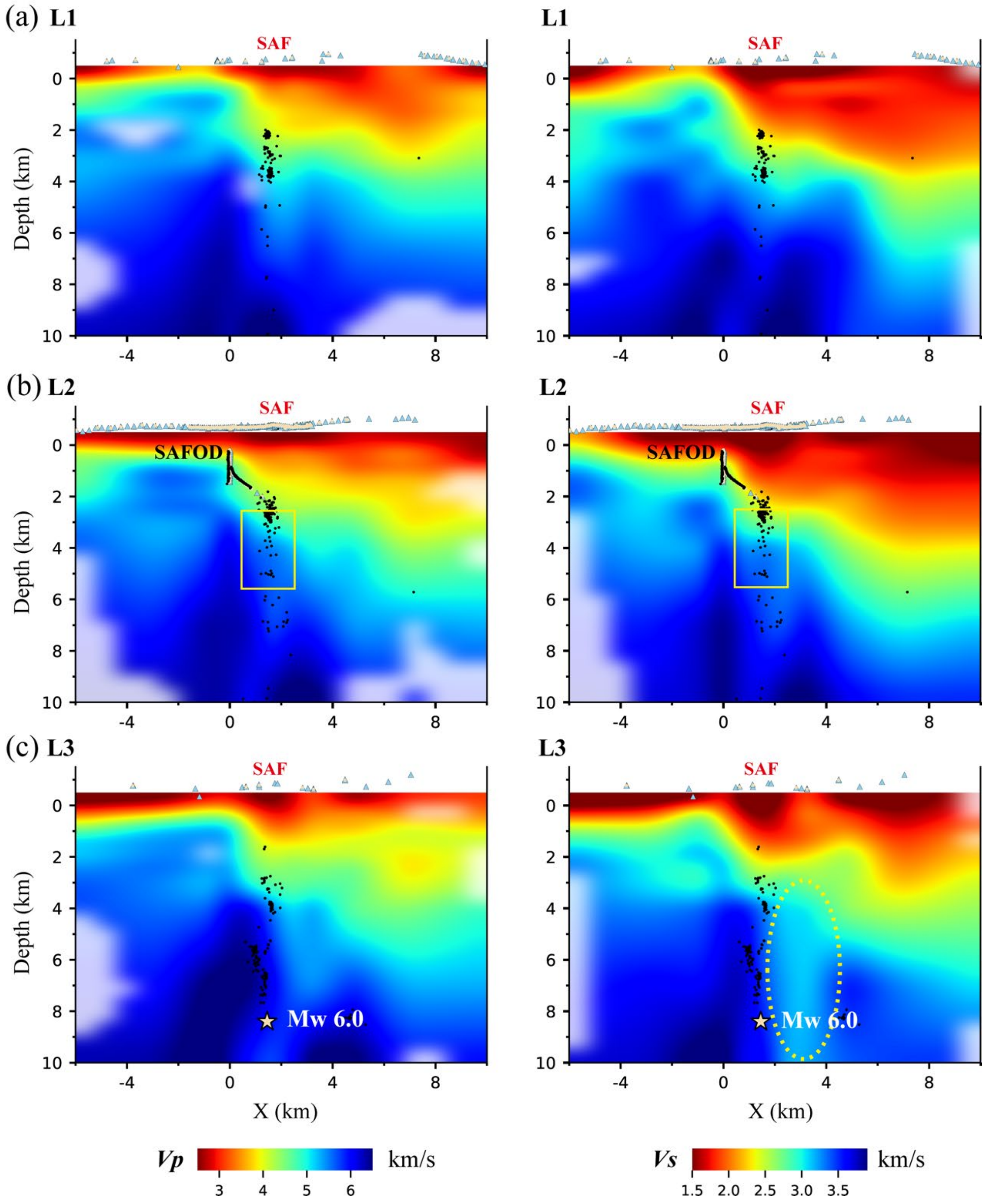

Figure 12. Velocity profiles across the SAF. Mean $V_p$ (left panels) and $V_s$ (right panels) are shown on vertical sections across the SAF (L1-L3 shown in Fig. 10). Relocated earthquakes occurring within ±1 km from the profile are shown, and yellow stars represent the Mw 6.0 1966 epicenter. Triangles represent used stations. The white-shaded areas represent regions with high standard deviations ( $V_p$ std > 0.34 and $V_s$ std > 0.23), which indicate poor data constraints.

### 4.3 Comparison with standard double difference tomography

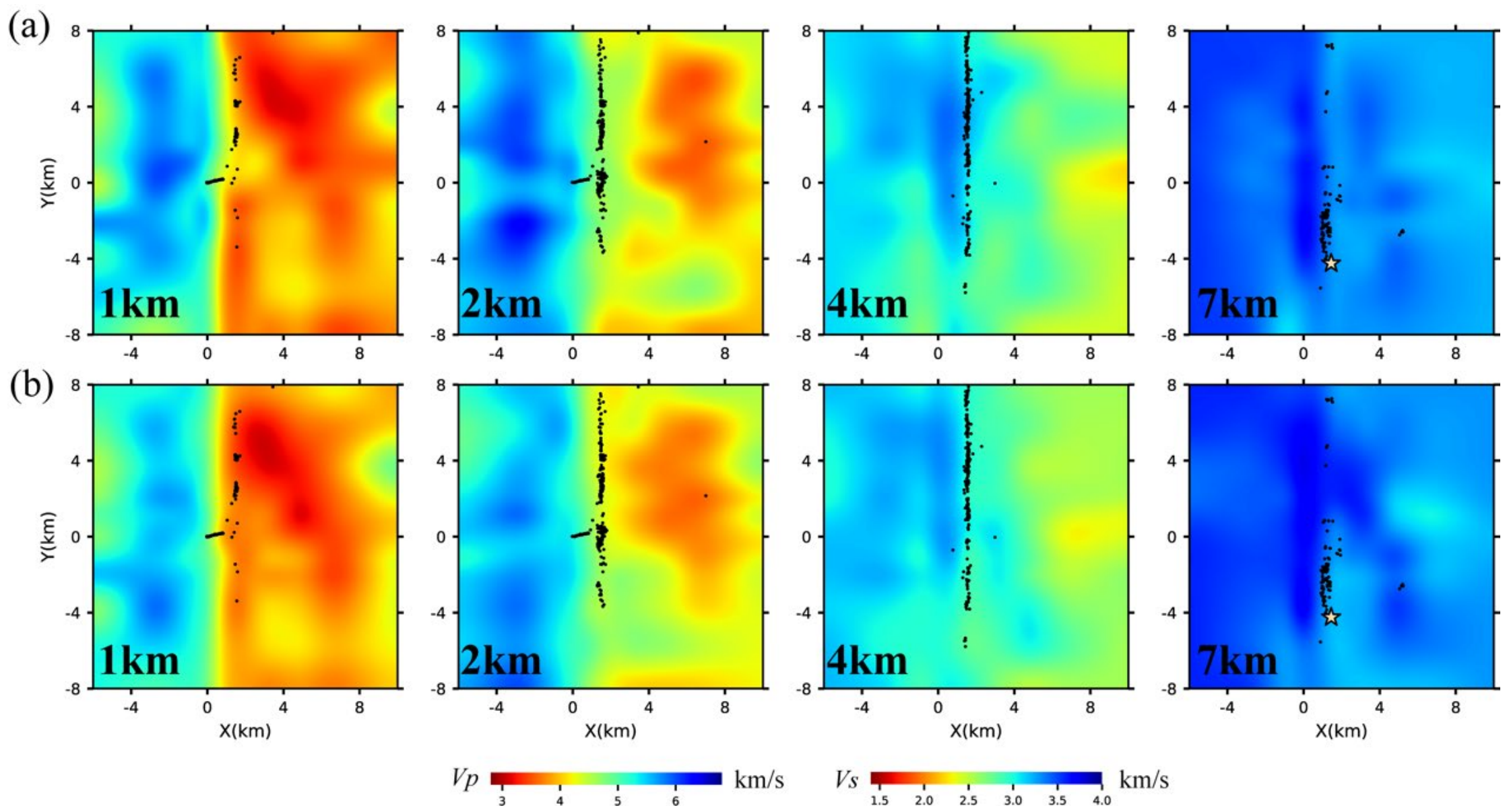

Figure 13. Horizontal slices of the Vp and Vs models at different depths inverted by tomoDD.

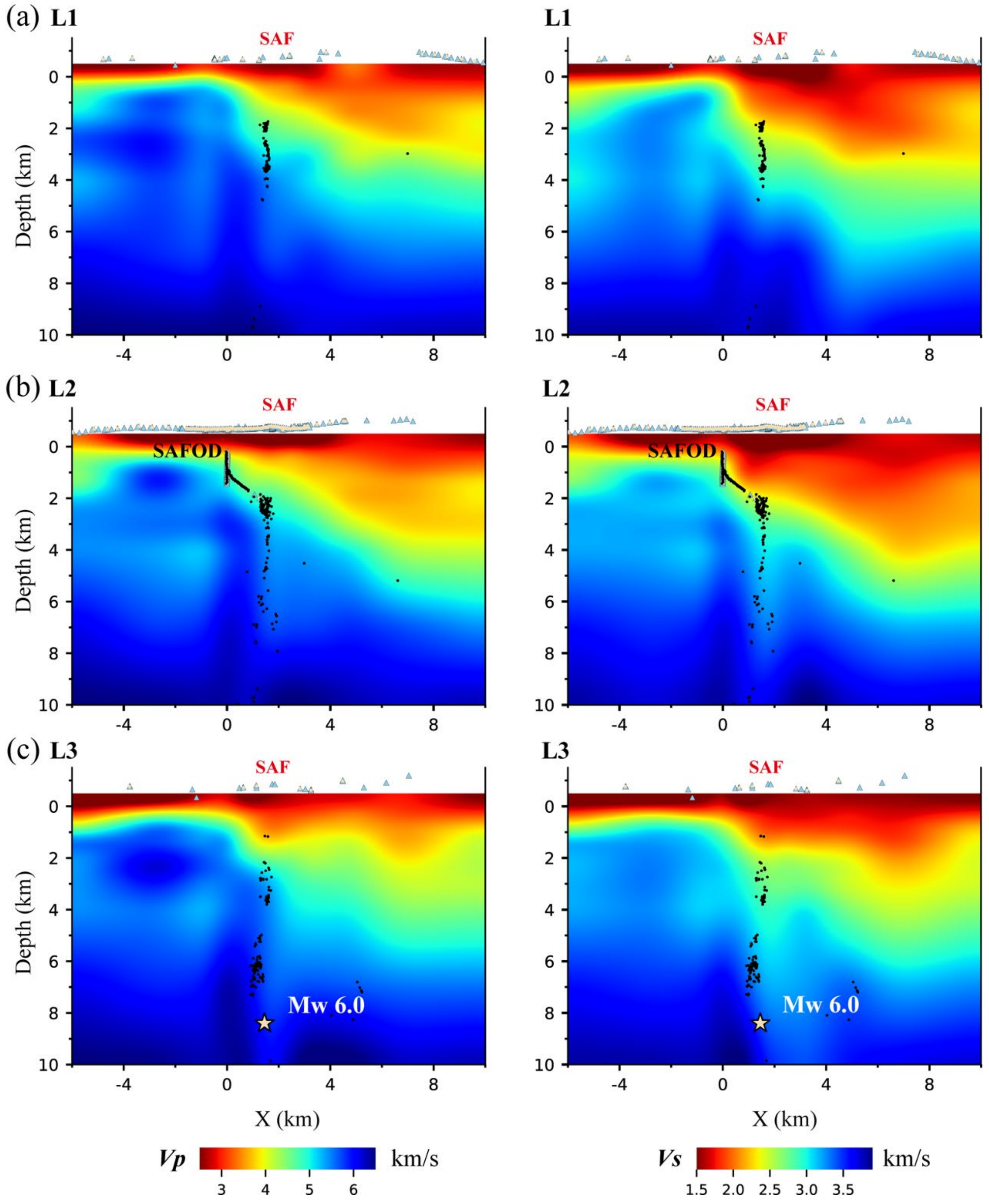

Figure 14. Velocity sections across the SAF by tomoDD. Keys are the same as in Fig. 12.

We conducted a set of inversions using the traditional DD tomography algorithm based on the same dataset and model parameterization. Detailed inversion procedures are outlined in Appendix B. The results indicate that both methods reveal similar velocity structures, notably the distinct velocity contrast across the San Andreas Fault (Figs. 13 and 14). However, the traditional method yields smoother velocity models with a reduced velocity contrast. As discussed in section 3.2, this method incorporates regularization constraints to enhance inversion stability, assuming "simple" and "smooth" subsurface structures. Consequently, this approach may suppress anomaly magnitudes and eliminate localized small anomalies. For example, the low velocity anomaly in Fig 12 (denoted by yellow dashed ellipse) is smeared out in the results obtained using traditional double difference tomography (Fig. 14) .

In terms of data residuals, the variational DD tomography method shows a better fit to the actual data than the traditional method. The root mean square (RMS) residuals decrease from an initial value of 0.1129 s to 0.0529 s for variational DD tomography and to 0.0734 s for the traditional tomography (tomoDD). This indicates that variational DD tomography produces a model that matches more closely the observed data.

## 5 DISCUSSION

In this study, we combined traditional DD tomography with a variational method (sSVGD) to develop an efficient 3D Bayesian body-wave tomography technique. The variational inference DD tomography approach uses gradient descent to minimize the difference between target and estimated distributions, generating a posterior probability distribution for model parameters. Synthetic tests indicate that sSVGD can accurately approximate the posterior probability density function.

To enhance computational efficiency, we parallelized the ray tracing process in this study by distributing the computation across multiple cores. For large datasets, further improvements can be achieved by using random mini-batches (Liu & Wang, 2016). In the synthetic tests in section 3, the relocation test took 2 hours on a personal computer with 20 Inter Core i7-12700K cores, while the joint inversion test required 20 hours. In section 4, the real data application took about 20 hours using a server with 80 Intel Xeon Gold 6448H cores. This demonstrates that the method can be effectively applied to real data in terms of computational performance.

We found that tomoDD inevitably exhibits a smearing effect due to regularization applied to the inversion system. This regularization often involves a subjective perception of the subsurface structure (Bodin & Sambridge, 2009). In variational DD tomography, the smearing effect is significantly reduced due to the absence of regularization constraints, and the magnitudes of anomalous bodies are better recovered. However, we notice that insufficient sampling can lead to biases in areas with poor data coverage as observed in several previous studies (Zhang & Curtis, 2023; Zhang *et al.*, 2023; Zhao *et al.*, 2022). This bias can be mitigated by increasing the number of initial particles or the number of iterations.

In this study, we used a fixed grid to parameterize the model, which may limit the ability to describe extreme cases such as regions with sharp velocity changes. If the grid has too few cells, it could lead to underfitting, while overly fine grids might result in inadequate data coverage. For example, in real data inversion we set a very fine grid near the fault zone, leading to increased uncertainty in those areas. To improve flexibility of model parameterization, more complex methods can be employed, such as Voronoi diagrams (Bodin & Sambridge, 2009; Sambridge *et al.*, 1995) and wavelet representations (Fang *et al.*, 2015; Zhang & Zhang, 2015). Additionally, using multiple regular grid parameterization may yield better or at least comparable results (Tong, 2021).

In linearized inversion methods, the checkerboard test is usually used to evaluate spatial resolution, and to provide quick and intuitive understanding of the reliability of the obtained model. However, according to our tests in Section 3.2, checkerboard tests cannot accurately determine the reliability of observed anomalies, and consequently these results become difficult to interpret in practice. By contrast, variational DD tomography estimates the posterior probability distribution, and can therefore provide accurate uncertainty estimates of velocity structures. The method therefore provides an important tool to assess reliability of obtained models and to improve interpretation of the results.

In this study, we used independent uniform distributions to parameterize each cell,

which might not be effective for large-scale inverse problems (Curtis & Lomax, 2001; Earp & Curtis, 2020). To address these issues, more robust prior information or structure-based priors can be used to ensure exploration close to the global optimum and to improve computational efficiency (Caers, 2018; Nawaz & Curtis, 2018; Walker & Curtis, 2014). In the real data inversion, we used a reliable three-dimensional velocity model as the prior information. If there is no suitable 3D model for the study area, we can use more efficient and stable techniques like tomoDD (Zhang & Thurber, 2003) to obtain relatively accurate prior information, which can improve inversion efficiency. On the other hand, traditional linear inversion methods and FWI have a high dependency on the initial model (Ao *et al.*, 2015; Datta & Sen, 2016; Kissling *et al.*, 1994), whereas sSVGD has lower requirements (Zhang *et al.*, 2023). Therefore, in cases with little or no reference information, using a broader prior distribution in variational DD tomography can provide a relatively accurate initial velocity model for linearized methods.

## 6 CONCLUSION

In this study, we integrate variational inference (sSVGD) with classical DD tomography to develop a novel variational DD tomography method. We applied the method to both synthetic and real data tomographic problems and compared the results with those obtained using traditional approaches. The results show that the new method can solve 3D tomographic problems efficiently and provide accurate uncertainty estimates. Compared to the traditional checkerboard test, which can misrepresent the reliability of velocity anomalies, the variational DD tomography offers more accurate uncertainty estimates. In addition, the method can eliminate the artifacts that are caused by regularization in traditional methods. In the real data application around the SAFOD site, we observed sharper velocity contrasts in fault zones compared to standard DD tomography, along with reliable uncertainty estimates for velocity and seismic sources. We thus conclude that variational DD tomography provides a better tool for imaging subsurface structures.

## ACKNOWLEDGMENTS

The study is funded by National Natural Science Foundation of China under grant 42230101 and 42204055.

## DATA AVAILABILITY STATEMENTS

The code underlying this paper can be downloaded from GitHub (https://github.com/xin2zhang/VIP).

## REFERENCES

Agata, R., Shiraishi, K. & Fujie, G., 2023, Bayesian seismic tomography based on velocity-space Stein variational gradient descent for physics-informed neural network, in *IEEE Transactions on Geoscience and Remote Sensing*, Vol. **61**, pp. 1-17.

Aki, K. & Lee, W., 1976, Determination of three‐dimensional velocity anomalies under a seismic array using first P arrival times from local earthquakes: 1. A homogeneous initial model, *J. Geophys. Res.*, **81**(23), 4381-4399.

Aleardi, M., Vinciguerra, A. & Hojat, A., 2021, A geostatistical Markov chain Monte Carlo inversion algorithm for electrical resistivity tomography, *Near Surf. Geophys.*, **19**(1), 7-26.

Andrieu, C. & Thoms, J., 2008, A tutorial on adaptive MCMC, *Stat. Comput.* , **18**, 343-373.

Ao, R.-D., Dong, L.-G. & Chi, B.-X., 2015, Source-independent envelope-based FWI to build an initial model, *Chin. J. Geophys.*, **58**(6), 1998-2010.

Ba, J., Erdogdu, M. A., Ghassemi, M., Sun, S., Suzuki, T., Wu, D. & Zhang, T., Understanding the variance collapse of SVGD in high dimensions, *in* Proceedings International Conference on Learning Representations2021.

Bishop, C., 2006, Pattern recognition and machine learning, *Springer*, **2**, 5-43.

Blatter, D., Key, K., Ray, A., Gustafson, C. & Evans, R., 2019, Bayesian joint inversion of controlled source electromagnetic and magnetotelluric data to image freshwater aquifer offshore New Jersey, *Geophys. J. Int.*, **218**(3), 1822-1837.

Blei, D. M. & Jordan, M. I., 2006, Variational inference for Dirichlet process mixtures, *Bayesian Anal.*, **1**(1), 121-143.

Blei, D. M., Kucukelbir, A. & McAuliffe, J. D., 2017, Variational inference: A review for statisticians, *J. Am. Stat. Assoc.*, **112**(518), 859-877.

Bodin, T. & Sambridge, M., 2009, Seismic tomography with the reversible jump algorithm, *Geophys. J. Int.*, **178**(3), 1411-1436.

Bosch, M., Meza, R., Jiménez, R. & Hönig, A., 2006, Joint gravity and magnetic inversion in 3D using Monte Carlo methods, *Geophysics*, **71**(4), G153-G156.

Box, G. E. & Tiao, G. C., 2011, Bayesian inference in statistical analysis, John Wiley & Sons.

Caers, J., 2018, Bayesianism in the Geosciences, *Handb. Math. Geosci.*, 527-566.

Chen, C., Carlson, D., Gan, Z., Li, C. & Carin, L., 2016, Bridging the gap between stochastic gradient MCMC and stochastic optimization, *in* Proceedings Artificial Intelligence and Statistics2016, PMLR, pp. 1051-1060.

Chen, T., Fox, E. & Guestrin, C., 2014, Stochastic gradient hamiltonian monte carlo, *in* Proceedings International conference on machine learning2014, PMLR, pp. 1683-1691.

Curtis, A. & Lomax, A., 2001, Prior information, sampling distributions, and the curse of dimensionality, *Geophysics*, **66**(2), 372-378.

Datta, D. & Sen, M. K., 2016, Estimating a starting model for full-waveform inversion using a global optimization method, *Geophysics*, **81**(4), R211-R223.

Dixit, M. M., Kumar, S., Catchings, R. D., Suman, K., Sarkar, D. & Sen, M., 2014, Seismicity, faulting, and structure of the Koyna - Warna seismic region, Western India from local earthquake tomography and hypocenter locations, *J. Geophys. Res. Solid Earth*, **119**(8), 6372-6398.

Durmus, A., Majewski, S. & Miasojedow, B., 2019, Analysis of Langevin Monte Carlo via convex optimization, *J. Mach. Learn. Res.*, **20**(1), 2666-2711.

Earp, S. & Curtis, A., 2020, Probabilistic neural network-based 2D travel-time tomography, *Neural Comput. Appl.*, **32**(22), 17077-17095.

El Ghaoui, L. & Lebret, H., 1997, Robust solutions to least-squares problems with uncertain data, *SIAM J. Matrix Anal. Appl.*, **18**(4), 1035-1064.

Fang, H., Yao, H., Zhang, H., Huang, Y.-C. & van der Hilst, R. D., 2015, Direct inversion of surface wave dispersion for three-dimensional shallow crustal structure based on ray tracing: methodology and application, *Geophys. J. Int.*, **201**(3), 1251-1263.

Fichtner, A., Zunino, A. & Gebraad, L., 2019, Hamiltonian Monte Carlo solution of tomographic inverse problems, *Geophys. J. Int.*, **216**(2), 1344-1363.

Galetti, E., Curtis, A., Meles, G. A. & Baptie, B., 2015, Uncertainty loops in travel-time tomography from nonlinear wave physics, *Phys. Rev. Lett.*, **114**(14), 148501.

Gallego, V. & Insua, D. R., 2018, Stochastic gradient MCMC with repulsive forces, preprint (arXiv:1812.00071)

Girolami, M. & Calderhead, B., 2011, Riemann manifold langevin and hamiltonian monte carlo methods, *J. R. Stat. Soc. Ser. B Stat. Methodol.*, **73**(2), 123-214.

Green, P. J. & Hastie, D. I., 2009, Reversible jump MCMC, *Genetics*, **155**(3), 1391-1403.

Guo, P., Visser, G. & Saygin, E., 2020, Bayesian trans-dimensional full waveform inversion: synthetic and field data application, *Geophys. J. Int.*, **222**(1), 610-627.

Hansen, P. C., 1999, The L-curve and its use in the numerical treatment of inverse problems, 119-142.

Izquierdo, K., Lekić, V. & Montési, L. G., 2020, A Bayesian approach to infer interior mass anomalies from the gravity data of celestial bodies, *Geophys. J. Int.*, **220**(3), 1687-1699.

Kissling, E., Ellsworth, W., Eberhart‐Phillips, D. & Kradolfer, U., 1994, Initial reference models in local earthquake tomography, *J. Geophys. Res. Solid Earth*, **99**(B10), 19635-19646.

Kucukelbir, A., Tran, D., Ranganath, R., Gelman, A. & Blei, D. M., 2017, Automatic differentiation variational inference, *J. Mach. Learn. Res.*, **18**(14), 1-45.

Kullback, S. & Leibler, R. A., 1951, On information and sufficiency, *Ann. Math. Stat.*, **22**(1), 79-86.

Leviyev, A., Chen, J., Wang, Y., Ghattas, O. & Zimmerman, A., 2022, A stochastic Stein Variational Newton method, preprint (arXiv:2204.09039).

Li, G., Niu, F., Yang, Y. & Tao, K., 2019, Joint inversion of Rayleigh wave phase velocity, particle motion, and teleseismic body wave data for sedimentary structures, *Geophys. Res. Lett.*, **46**(12), 6469-6478.

Li, Z., Tian, B., Liu, S. & Yang, J., 2013, Asperity of the 2013 Lushan earthquake in the eastern margin of Tibetan Plateau from seismic tomography and aftershock relocation, *Geophys. J. Int.*, **195**(3), 2016-2022.

Liu, Q., 2017, Stein variational gradient descent as gradient flow, *in Advances In Neural Information Processing Systems*, **30**(

Liu, Q. & Wang, D., 2016, Stein variational gradient descent: A general purpose bayesian inference algorithm, in *Advances In Neural Information Processing Systems*, pp. 2378-2386.

Liu, Y., Yao, H., Zhang, H. & Fang, H., 2021, The community velocity model V. 1.0 of southwest China, constructed from joint body‐ and surface‐ wave travel‐ time tomography, *Seismol. Soc. Am.*, **92**(5), 2972-2987.

Lomax, A. & Curtis, Fast, probabilistic earthquake location in 3D models using oct-tree importance sampling, *Geophys. Res. Abstr,* 3, pp. 955.

Luo, X., 2010, Constraining the shape of a gravity anomalous body using reversible jump Markov chain Monte Carlo, *Geophys. J. Int.*, **180**(3), 1067-1079.

Malinverno, A., 2002, Parsimonious Bayesian Markov chain Monte Carlo inversion in a nonlinear geophysical problem, *Geophys. J. Int.*, **151**(3), 675-688.

Mandolesi, E., Ogaya, X., Campanyà, J. & Agostinetti, N. P., 2018, A reversible-jump Markov chain Monte Carlo algorithm for 1D inversion of magnetotelluric data, *Comput. Geosci.*, **113**, 94-105.

Martin, J., Wilcox, L. C., Burstedde, C. & Ghattas, O., 2012, A stochastic Newton MCMC method for large-scale statistical inverse problems with application to seismic inversion, *SIAM J. Sci. Comput.*, **34**(3), A1460-A1487.

Meng, L., McGarr, A., Zhou, L. & Zang, Y., 2019, An investigation of seismicity induced by hydraulic fracturing in the Sichuan Basin of China based on data from a temporary seismic network, *Bull. Seismol. Soc. Am.*, **109**(1), 348-357.

Nawaz, M. A. & Curtis, A., 2018, Variational Bayesian inversion (VBI) of quasi-localized seismic attributes for the spatial distribution of geological facies, *Geophys. J. Int.*, **214**(2), 845-875.

-, 2019, Rapid discriminative variational Bayesian inversion of geophysical data for the spatial distribution of geological properties, *J. Geophys. Res. Solid Earth*, **124**(6), 5867-5887.

Neal, R. M., 1993, Probabilistic inference using Markov chain Monte Carlo methods.

Nemeth, C. & Fearnhead, P., 2021, Stochastic gradient markov chain monte carlo, *J. Am. Stat. Assoc.*, **116**(533), 433-450.

Pesicek, J., Thurber, C., Zhang, H., DeShon, H., Engdahl, E. & Widiyantoro, S., 2010, Teleseismic double‐difference relocation of earthquakes along the Sumatra‐Andaman subduction zone using a 3‐D model, *J. Geophys. Res. Solid Earth*, **115**, B10303.

Piana Agostinetti, N., Giacomuzzi, G. & Chiarabba, C., 2020, Across‐fault velocity gradients and slip behavior of the San Andreas Fault near Parkfield, *Geophys. Res. Lett.*, **47**(1), e2019GL084480.

Piana Agostinetti, N., Giacomuzzi, G. & Malinverno, A., 2015, Local three-dimensional earthquake tomography by trans-dimensional Monte Carlo sampling, *Geophys. J. Int.*, **201**(3), 1598-1617.

Ramgraber, M., Weatherl, R., Blumensaat, F. & Schirmer, M., 2021, Non‐Gaussian Parameter Inference for Hydrogeological Models Using Stein Variational Gradient Descent, *Water Resour. Res.*, **57**(4), e2020WR029339.

Ramirez, A. L., Nitao, J. J., Hanley, W. G., Aines, R., Glaser, R. E., Sengupta, S. K., Dyer, K. M., Hickling, T. L. & Daily, W. D., 2005, Stochastic inversion of electrical resistivity changes using a Markov Chain Monte Carlo approach, *J. Geophys. Res. Solid Earth*, **110**, B02101.

Rawlinson, N. & Spakman, W., 2016, On the use of sensitivity tests in seismic tomography, *Geophys. J. Int.*, **205**(2), 1221-1243.

Sambridge, M., Braun, J. & McQueen, H., 1995, Geophysical parametrization and interpolation of irregular data using natural neighbours, *Geophys. J. Int.*, **122**(3), 837-857.

Smith, J. D., Ross, Z. E., Azizzadenesheli, K. & Muir, J. B., 2022, HypoSVI: Hypocentre inversion with Stein variational inference and physics informed neural networks, *Geophys. J. Int.*, **228**(1), 698-710.

Tarantola, A., 2005, Inverse problem theory and methods for model parameter estimation, SIAM.

Thurber, C., Zhang, H., Brocher, T. & Langenheim, V., 2009, Regional three‐dimensional seismic velocity model of the crust and uppermost mantle of northern California, *J. Geophys. Res. Solid Earth*, **114**, B01304.

Thurber, C., Zhang, H., Waldhauser, F., Hardebeck, J., Michael, A. & Eberhart-Phillips, D., 2006, Three-dimensional compressional wavespeed model, earthquake relocations, and focal mechanisms for the Parkfield, California, region, *Bull. Seismol. Soc. Am.*, **96**(4B), S38-S49.

Tong, P., 2021, Adjoint - state traveltime tomography: Eikonal equation - based methods and application to the Anza area in southern California, *J. Geophys. Res. Solid Earth*, **126**(5), e2021JB021818.

Tromp, J., Tape, C. & Liu, Q. Y., 2005, Seismic tomography, adjoint methods, time reversal and banana-doughnut kernels, *Geophys. J. Int.*, **160**(1), 195-216.

Um, J. & Thurber, C., 1987, A fast algorithm for two-point seismic ray tracing, *Bull. Seismol. Soc. Am.*, **77**(3), 972-986.

Waldhauser, F., 2001, HypoDD-A program to compute double-difference hypocenter locations, 2331-1258.

Walker, M. & Curtis, A., 2014, Varying prior information in Bayesian inversion, *Inverse Probl.*, **30**(6), 065002.

Welling, M. & Teh, Y. W., Bayesian learning via stochastic gradient Langevin dynamics, *in* Proceedings Proceedings of the 28th international conference on machine learning (ICML-11)2011, Citeseer, pp. 681-688.

Westman, E., Luxbacher, K., Schafrik, S., Swanson, P. & Zhang, H., Time-lapse passive seismic velocity tomography of longwall coal mines: a comparison of methods, in *Proceedings ARMA US Rock Mechanics/Geomechanics Symposium2012*, ARMA, pp. ARMA-2012-2589.

Yao, H., Ren, Z., Tang, J., Guo, R. & Yan, J., 2023, Trans-dimensional Bayesian joint inversion of magnetotelluric and geomagnetic depth sounding responses to constrain mantle electrical discontinuities, *Geophys. J. Int.*, **233**(3), 1821-1846.

Zhang, C. & Chen, T., 2022, Bayesian slip inversion with automatic differentiation variational inference, *Geophys. J. Int.*, **229**(1), 546-565.

Zhang, H., Liu, Y., Thurber, C. & Roecker, S., 2007, Three - dimensional shear - wave splitting tomography in the Parkfield, California, region, *Geophys. Res. Lett.*, **34**(24),

Zhang, H., Sarkar, S., Toksöz, M. N., Kuleli, H. S. & Al-Kindy, F., 2009a, Passive seismic tomography using induced seismicity at a petroleum field in Oman, *Geophysics*, **74**(6), WCB57-WCB69.

Zhang, H., Thurber, C. & Bedrosian, P., 2009b, Joint inversion for vp, vs, and vp/vs at SAFOD, Parkfield, California, *Geochem. Geophys. Geosyst.*, **10**(11),

Zhang, H., Thurber, C. H., Shelly, D., Ide, S., Beroza, G. C. & Hasegawa, A., 2004, High-resolution subducting-slab structure beneath northern Honshu, Japan, revealed by double-difference tomography, *Geology*, **32**(4), 361-364.

Zhang, H. J. & Thurber, C. H., 2003, Double-difference tomography: The method and its application to the Hayward Fault, California, *Bull. Seismol. Soc. Am.*, **93**(5), 1875-1889.

Zhang, X. & Curtis, A., 2020a, Seismic tomography using variational inference methods, *J. Geophys. Res. Solid Earth*, **125**(4), e2019JB018589.

Zhang, X. & Curtis, A., 2020b, Variational full-waveform inversion, *Geophys. J. Int.*, **222**(1), 406-411.
Zhang, X. & Curtis, A., 2021, Bayesian full-waveform inversion with realistic priors, *Geophysics*, **86**(5), A45-A49.
Zhang, X. & Curtis, A., 2023, VIP--Variational Inversion Package with example implementations of Bayesian tomographic imaging, preprint (arXiv:2310.13325)
Zhang, X., Curtis, A., Galetti, E. & De Ridder, S., 2018, 3-D Monte Carlo surface wave tomography, *Geophys. J. Int.*, **215**(3), 1644-1658.
Zhang, X., Hansteen, F., Curtis, A. & De Ridder, S., 2020a, 1 - D, 2 - D, and 3 - D Monte Carlo Ambient Noise Tomography Using a Dense Passive Seismic Array Installed on the North Sea Seabed, *J. Geophys. Res. Solid Earth*, **125**(2), e2019JB018552.
Zhang, X., Lomas, A., Zhou, M., Zheng, Y. & Curtis, A., 2023, 3-D Bayesian variational full waveform inversion, *Geophys. J. Int.*, **234**(1), 546-561.
Zhang, X., Roy, C., Curtis, A., Nowacki, A. & Baptie, B., 2020b, Imaging the subsurface using induced seismicity and ambient noise: 3-D tomographic Monte Carlo joint inversion of earthquake body wave traveltimes and surface wave dispersion, *Geophys. J. Int.*, **222**(3), 1639-1655.
Zhang, X. & Zhang, H., 2015, Wavelet - based time - dependent travel time tomography method and its application in imaging the Etna volcano in Italy, *J. Geophys. Res. Solid Earth*, **120**(10), 7068-7084.
Zhao, X., Curtis, A. & Zhang, X., 2022, Bayesian seismic tomography using normalizing flows, *Geophys. J. Int.*, **228**(1), 213-239.
Zhao, Z. & Sen, M. K., 2019, A gradient based MCMC method for FWI and uncertainty analysis, in *Proceedings SEG International Exposition and Annual Meeting2019*, SEG, pp. D043S130R001.
Zhao, Z. & Sen, M. K., 2019, 2021, A gradient-based Markov chain Monte Carlo method for full-waveform inversion and uncertainty analysis, *Geophysics*, **86**(1), R15-R30.
Zhu, H., 2018, Seismogram registration via Markov chain Monte Carlo optimization and its applications in full waveform inversion, *Geophys. J. Int.*, **212**(2), 976-987.

## APPENDIX A: VELOCITY UNCERTAINTY IN PARKFIELD ZONE

Fig. A1 shows the standard deviation in the horizontal slice of the inversion results for the Parkfield area, while Figs. A2 present the standard deviation at three profiles along L1-L3. The prior standard deviation for $V_p$ and $V_s$ are approximately 0.34 km/s and 0.23 km/s, respectively. After inversion, the standard deviation in most areas dropped below 0.34 km/s and 0.23 km/s. This indicates that these regions are better constrained by the data, as the posterior standard deviation is significantly lower than the prior standard deviation. However, around the fault rupture zone (X=0 km) where the velocity changes abruptly, the standard deviation shows high uncertainty. This has also been observed in several previous studies (Galetti *et al.*, 2015; Zhang *et al.*, 2018) which reflects the uncertainty of the location of velocity contrasts.

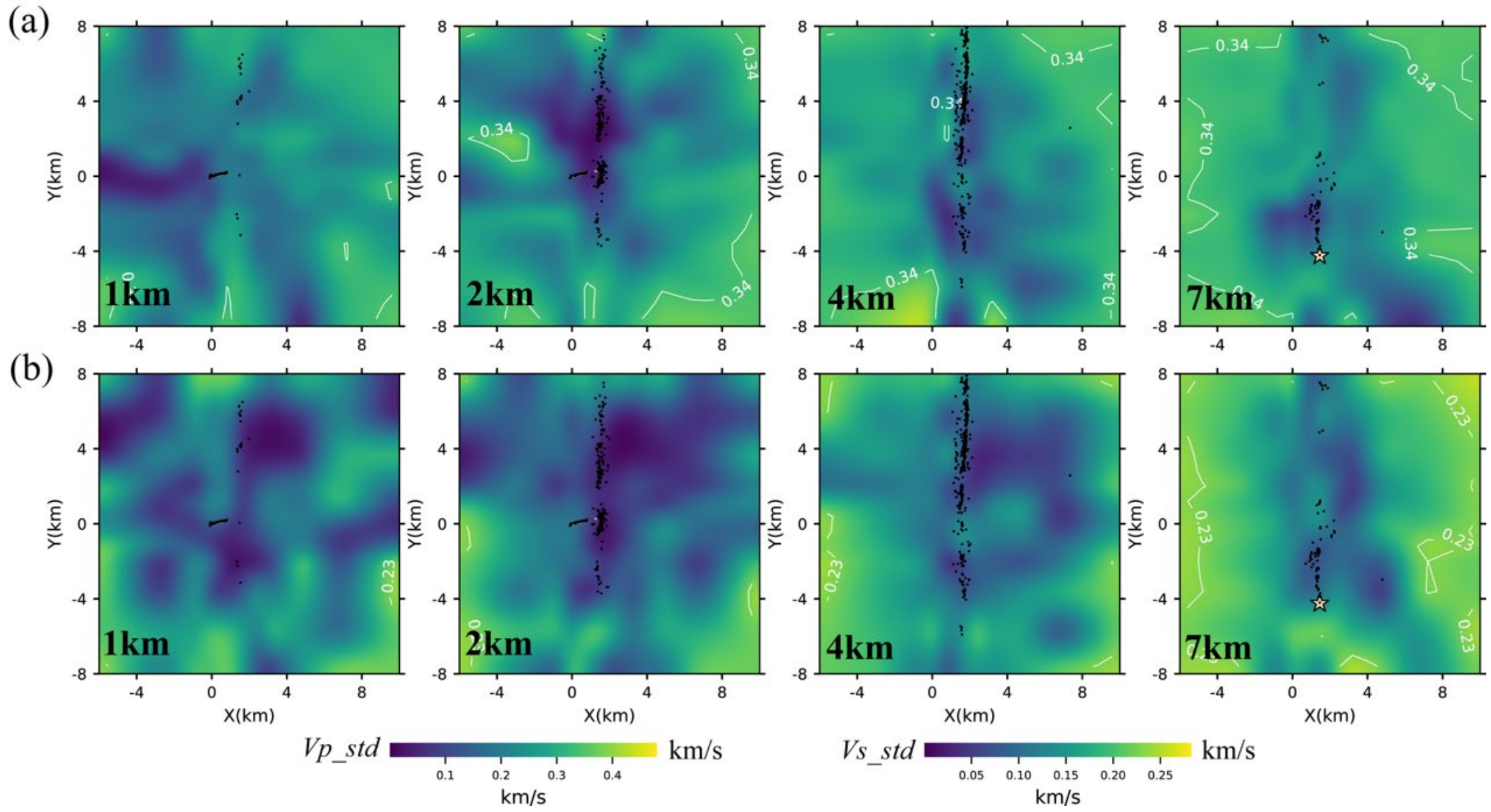

Figure A1. Horizontal sections of standard deviation of $V_p$ (top panels) and $V_s$ (lower panels) at 1, 2, 4, and 7km depths. Relocated earthquakes occurring within ±1 km from the layer are shown as black dots. Regions with posterior standard deviations below 0.34 and 0.23 for $V_p$ and $V_s$, respectively, are considered well-constrained by the data.

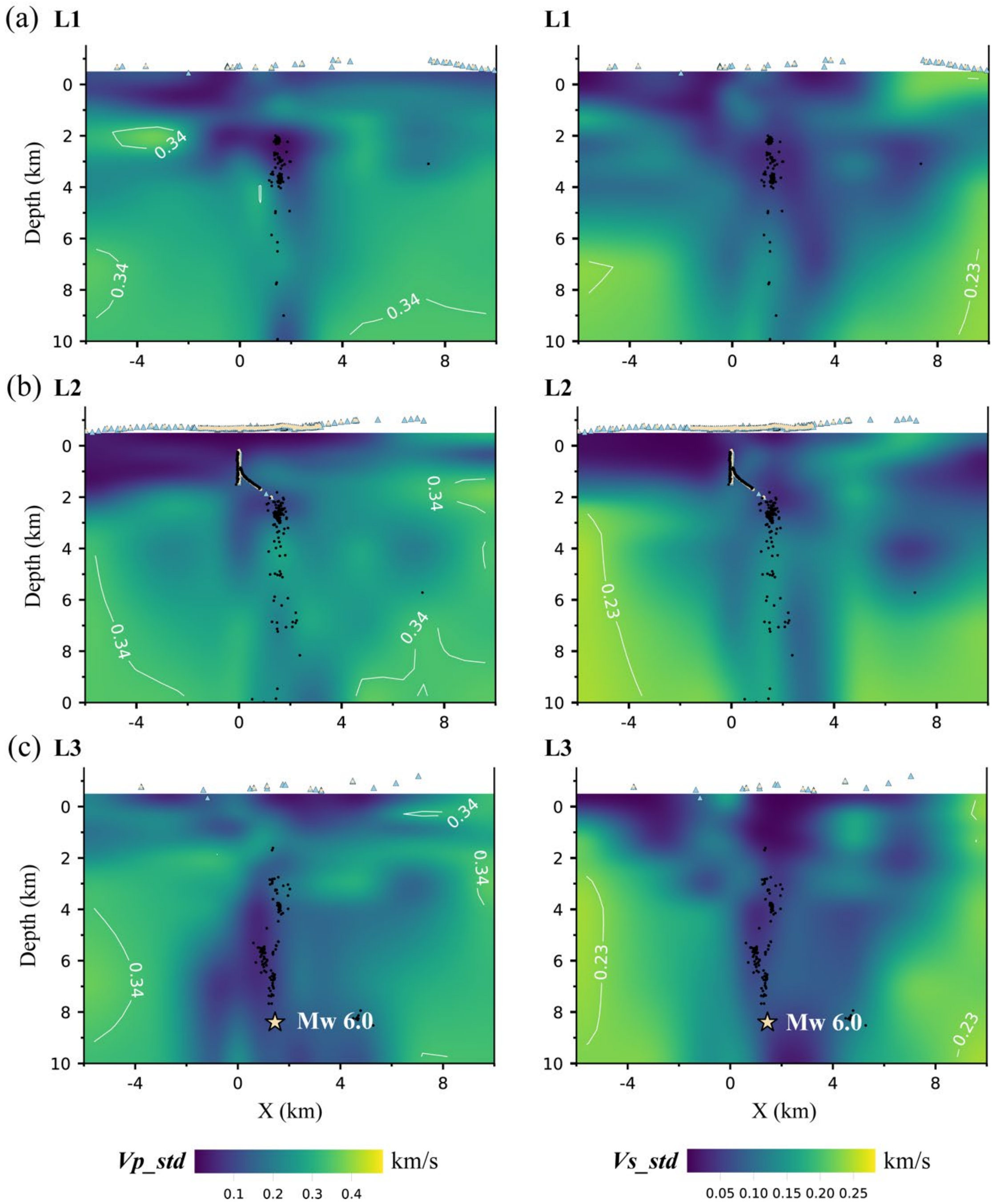

Figure A2. Vertical sections of standard deviation $V_p$ (left panels) and $V_s$ (right panels) across the SAF ( L1-L3 shown in Fig. 10). Relocated earthquakes occurring within ±1 km from the profile is shown, and yellow stars represent the Mw 6.0 1966 epicenter. Triangles represent used stations. Regions with posterior standard deviations below 0.34 and 0.23 for $V_p$ and $V_s$, respectively, are considered well-constrained by the data.

## APPENDIX B: TOMODD INVERSION DETAILS

To compare different methods, we also used tomodd to obtain velocity models. The inversion uses the same data and model parameterization as variational double difference tomography. The initial model is set as the average of the prior information used by the variational method, which is considered a fairly accurate 3D model. Fig. B1 shows the selection of regularization parameters. We chose the regularization parameters as the inflection point of the L-curve (damping=70, smoothing=5) to balance data constraints with model constraints.

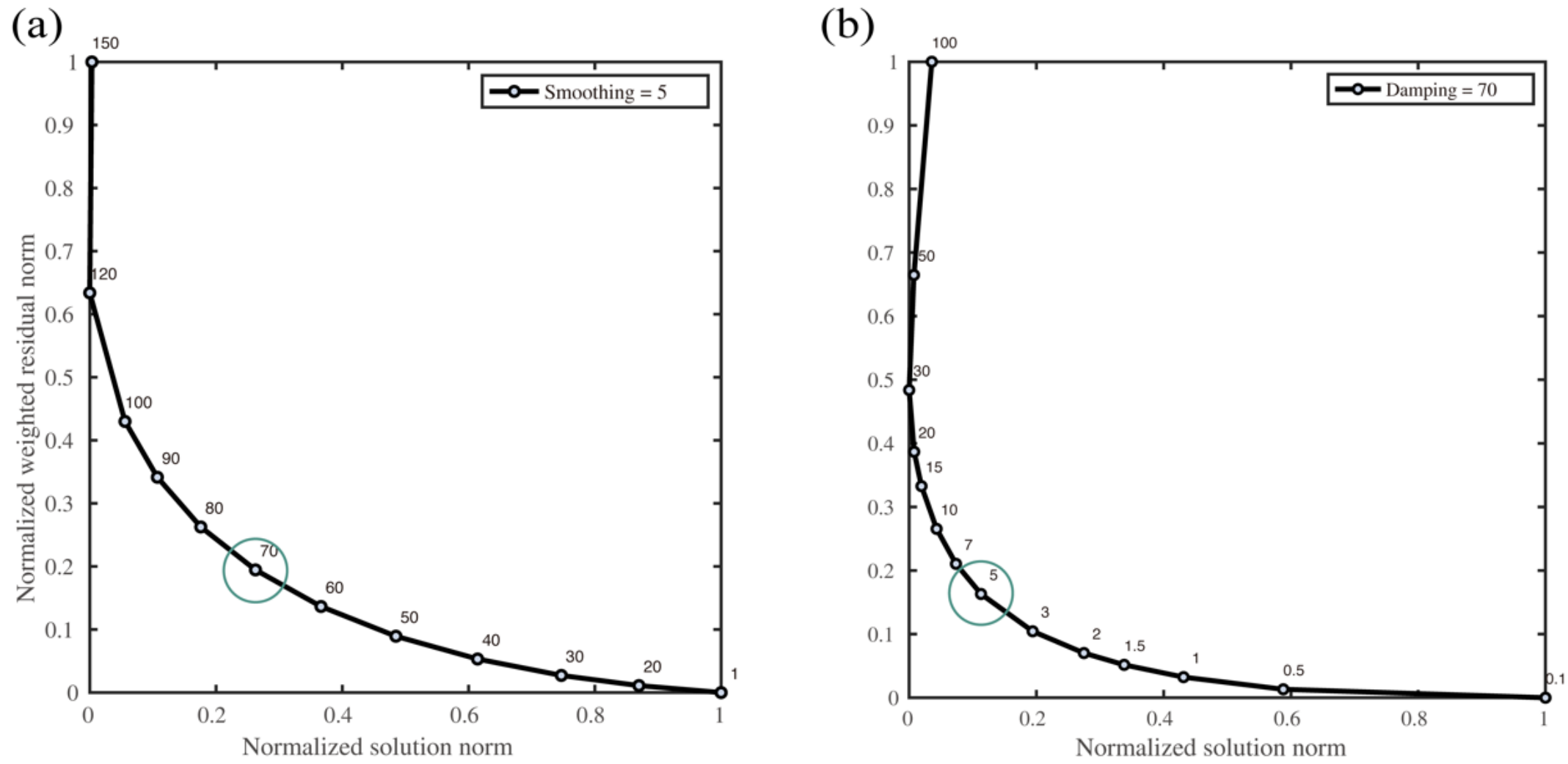

Figure B1. The L-Curve for the selection of optional smoothing (a) and damping (b). The green circle highlights the regularization parameters we chose: damping is set to 70 and smoothing to 5.